\documentclass[preprint,prc,showpacs,preprintnumbers,amsmath,amssymb,floatfix]{revtex4-1}

\usepackage{amsmath}
\usepackage{lipsum}
\usepackage{graphicx}
\usepackage{dcolumn}
\usepackage{bm}
\usepackage{ulem} 
\usepackage[usenames]{color}
\usepackage{epstopdf}
\usepackage{epsfig}
\usepackage{float}
\usepackage{subfigure}
\usepackage{multirow}
\usepackage[version=3]{mhchem}
\usepackage[backref]{hyperref}

\newcommand{\nc}{\newcommand}       
\nc{\vc}[1] {\mbox{\boldmath $#1$}} 
\nc{\del}       {\partial}              
\nc{\bra}       {\langle}               
\nc{\ket}       {\rangle}               
\nc{\bras}[1]   {\langle #1|}           
\nc{\kets}[1]   {|#1\rangle}            
\nc{\mapleft}[1]{           
	\smash{\mathop{\,          %
			\hbox to 1.5cm{\rightarrowfill}\, }\limits_{#1}}}
\nc{\beq}     {\begin{eqnarray}} \nc{\eeq}    {\end{eqnarray}}
\nc{\nn}      {\\\nonumber} \nc{\vs}      {\vspace{-0.275cm}}
\nc{\fra}    {\frac{1}{2}}
\nc{\mb}        {\mathbf}

\begin{document}
	
	\preprint{}
	
	\title{The hyperonic  star in relativistic mean-field model}
	
	\author{Kaixuan Huang}
	\affiliation{School of Physics, Nankai University, Tianjin 300071,  China}
	
	\author{Jinniu Hu}~\email{hujinniu@nankai.edu.cn}
	\affiliation{School of Physics, Nankai University, Tianjin 300071,  China}

	\author{Ying Zhang}~\email{yzhangjcnp@tju.edu.cn}
	\affiliation{Department of Physics, School of Science, Tianjin University, Tianjin 300072, China}

	\author{Hong Shen}~\email{songtc@nankai.edu.cn}
	\affiliation{School of Physics, Nankai University, Tianjin 300071,  China}

	\date{\today}
	\begin{abstract}
		The neutron star as a supernova remnant is attracting high attention recently due to the gravitation wave detection and precise measurements about its mass and radius. In the inner core region of the neutron star, the strangeness degrees of freedom, such as the hyperons, can be present, which is also named as a hyperonic star. In this work, the neutron star consisting of nucleons and leptons, and the hyperonic star including the hyperons will be reviewed in the framework of the relativistic mean-field (RMF) model. The popular non-linear and density-dependent RMF parametrizations in the market will be adopted to investigate the role of strangeness baryons in a hyperonic star on its mass, radius, tidal deformability, and other properties. Finally, the magnitudes of the coupling strengths between mesons and hyperons also will be discussed, which can generate the massive hyperonic star with present RMF parameter sets, when the vector coupling constants are strong.
	\end{abstract}
	
	
	
	\maketitle

\section{Introduction}
The star, whose mass is  in the range of  $8M_\odot\sim20M_\odot$, will undergo a supernova explosion at the end of its life. The remnant may form a very compact object mainly consisting of nucleons and leptons, i.e., neutron stars. Due to the strong magnetic field ($B\sim 10^{11}- 10^{15}$ G) and fast rotation of neutron star ($P\sim 10^{-3}-10$ s), it can emit the beam of electromagnetic radiation from the magnetic poles, which was detected from the earth as a pulsar. In the past 50 years, more than $3000$ pulsars were measured, whose typical masses are around $1\sim 2 M_\odot$ and radii are about $10$ km. Therefore, the neutron star should consist of very dense matter. Its central density is close to $5-8\rho_0$, where $\rho_0\sim0.16$ fm$^{-3}$ is the nuclear saturation density. It is difficult to investigate the properties of such supra-dense matter in the terrestrial nuclear laboratory until now~\cite{glendenning97,weber99,prakash97,weber05,chamel08,lattimer16,oezel16,oertel17,lia19,lia20}.

There were many great achievements in the observations of the neutron star in the past decade with the fast developments of astronomical techniques. Several massive neutron stars, whose masses are around $2M_\odot$, PSR J1614-2230 ($1.928\pm0.017M_\odot$)~\cite{demorest10,fonseca16,arzoumanian18}, PSR J0348+0432 ($2.01\pm0.04M_\odot$)~\cite{antoniadis13},  and PSR J0740+6620  ($2.08\pm0.07M_\odot$)~\cite{cromartie20,miller21} were measured within the relativistic Shapiro delay effect. In the August of 2017, the gravitational wave and the electromagnetic counterpart of a binary neutron-star merger were detected by LIGO/Virgo and other astronomical observatories for the first time as GW170817 event~\cite{abbott17,abbott18,abbott19}. It opens the door to the multi-messenger astronomy era. From the gravitational wave, the tidal deformability of the neutron star is extracted, which is strongly correlated with the structures of the neutron star. Five gravitational wave events which are relevant to the neutron star have been detected until now, i.e. GW170817, GW190425~\cite{abbott20a}, GW190814~\cite{abbott20b}, GW200105, and GW200115~\cite{abbott21a}. The masses of these compact objects are in a wide region from $0.86M_\odot$ to $2.67M_\odot$~\cite{abbott21b}. The dimensionless tidal deformability of $1.4M_\odot$ neutron star extracted from GW170817 was $190^{+390}_{-120}$~\cite{tews18,finstad18,fattoyev18}.

The structure of a static neutron star is separated into five regions. The outer layer is the atmosphere consisting of the atom and is very thin. In the next layer as the outer crust, the electrons of an atom are ionized and form the uniform Fermi gas, where the nuclei are immersed. With the density increasing, neutrons drip out from the neutron-rich nuclei and generate the neutron gas, which is called the inner crust. The inhomogeneous nuclear matter is simplified as the droplet, rod, slab, tube, and bubble, i.e. the pasta structure~\cite{ravenhall83,williams85,oyamatsu93,oyamatsu07,magierski02,avancini08,watanabe09,pais12,schneider14,caplan17}. When the nucleon density approaches $\rho_0/2$, heavy nuclei dissolve and the neutron star matter becomes homogeneous, which corresponds to the outer core of neutron star. It plays an essential role in determining the mass and radius of the neutron star~\cite{grill12,bao15,pais15,yang19,ji21}. In the inner core region, the baryons including the strangeness degree of freedom, such as $\Lambda,~\Sigma$, and $\Xi$ hyperons, will be present, when the Fermi energies of nucleons are larger than their chemical potentials, which is also called as a hyperonic star. Furthermore, the quarks in the baryons are deconfined and generate quark matter~\cite{glendenning92,schertler99,schertler00,steiner00,burgio02,menezes03,yang08,xu10,chen13,orsaria14,wu17,baym18,malerran19,wu19,annala20,ju21a,ju21b}.

{ In 1960, Ambartsumyan and Saakyan firstly discussed the appearances of various hyperons in the free dense matter through the chemical equilibrium conditions~\cite{ambartsumyan60}. The hyperons were investigated in neutron stars from the 1980s by Glendenning in the framework of relativistic mean-field (RMF) model, where the coupling strengths between the mesons and baryons were simply generated by the quark power counting rules~\cite{glendenning82,glendenning85}. Actually, the appearances of hyperons are strongly dependent on the hyperon-nucleon and hyperon-hyperon potentials, which can be extracted from the properties of various hypernuclei. Recently, an abundant single-$\Lambda$ hypernuclei were produced through ($K^-,~\pi^-$), ($\pi^+,~K^+$), and ($e,~e'K^+$) reactions from light to heavy mass regions~\cite{hashimoto06,feliciello15,gal16,hiyama18}. A few double-$\Lambda$ hypernuclei were discovered in the light nuclei~\cite{nakazawa10,ahn13} and the first deep bound state of single-$\Xi$ hypernuclei was confirmed in the $\Xi^-+^{14}\rm{N}$ ($_{\Xi^-}^{15}\rm{C}$) system~\cite{nakazawa15,hayakawa21,yoshimoto21}. The observations of these hypernuclei imply that the $\Lambda$-nucleon and $\Xi$-nucleon potentials in nuclear medium should be attractive around nuclear saturation density~\cite{friedman21,hu22}. }
 
{In the past 30 years, the RMF parametrizations about nucleon-nucleon and nucleon-hyperon interactions were largely improved through reproducing the ground-state properties of finite nuclei and above hypernuclei experimental data, that were adopted to investigate the hyperonic star~\cite{knorren95,schaffner96,shen02,cavagnoli11,weissenborn12,providencia13,weissenborn13,fortin17,zhang18,li18,fortin20,shangguan21,tu21,lopes22}. On the other hand, the strangeness degree of freedom in the neutron star was also discussed directly from the realistic baryon-baryon interactions with $ab~initio$ methods, such as Brueckner Hartree-Fock model~\cite{vidana00,zhou05,schulze06,schulze11,yamamoto13,yamamoto14,logoteta19},  auxiliary field diffusion Monte Carlo model~\cite{logoteta15}, relativistic Brueckner-Hartree-Fock model, and so on~\cite{sammarruca09,katayama14}. }

With the discoveries of two-times-solar-mass neutron stars,  a ``hyperon puzzle" was proposed, since the neutron star maximum mass will be reduced by about $20\%$ once the hyperons are self-consistently introduced in the nuclear many-body methods. Therefore, it is difficult to explain the existence of massive neutron stars with hyperons at the beginning. Many schemes were raised to solve such a puzzle. The repulsive contributions from three-body hyperon-nucleon-nucleon interaction may increase the hyperonic star mass~\cite{yamamoto14,logoteta15,logoteta19}.  A high-density $\sigma-$cut potential can generate $2M_\odot$ neutron stars with $\Lambda,~\Sigma$, and $\Xi$ hyperons in the relativistic mean-field model~\cite{zhang18}. Furthermore, the density-dependent RMF (DDRMF) model under the constraints of $\Lambda$ hypernuclei also can support the massive hyperonic star due to its strong repulsive components on the pressure at high density~\cite{long12}. Recently, the role of vector meson including the strange quark, $\phi$ was also discussed in the hyperonic star with various hyperons~\cite{tu21}.

In this paper, the hyperonic star in the framework of RMF and DDRMF models will be systematically calculated with the most popular parameterizations. The exchange mesons with strangeness quarks will be completely included in RMF framework. The coupling strengths between the mesons and baryons will be constrained with the latest hypernuclei experimental data. The effects of their magnitudes on the properties of hyperonic stars are discussed in the last part. This paper is arranged as follows. In Section 2, the formulas about the neutron star and hyperonic star with the RMF model are shown in detail. In Section 3, the parametrizations of the nonlinear RMF and DDRMF model are listed and the properties of neutron stars and hyperonic stars are calculated and discussed. In Section 4, the summary and conclusion will be given.

\section{THEORETICAL FRAMEWORK}
\subsection{The nonlinear Relativistic Mean-field Model}
The relativistic mean-field (RMF) model was constructed based on the one-boson-exchange picture for the nucleon-nucleon ($NN$) interaction. In the early version, there were only scalar ($\sigma$) and vector mesons ($\omega$), which present the attractive and repulsive components of $NN$ interaction, respectively~\cite{walecka74}. Then, the nonlinear terms of $\sigma$ meson were introduced to reduce the incompressibility of nuclear matter~\cite{boguta77}. The isovector meson, $\rho$ was introduced to correctly describe the neutron star matter~\cite{serot86}. The nonlinear term  of $\omega$ meson~\cite{tm1} and the coupling terms between the $\omega$ and $\rho$ mesons~\cite{toddrutel05} were also included to improve the high-density behaviors of nuclear matter and density-dependence of symmetry energy, respectively. Furthermore, a density-dependent RMF (DDRMF)  model~\cite{brockmann92} was proposed based on the developments of relativistic Brueckner-Hartree-Fock (RBHF) model. When the masses of exchanging mesons are regarded as infinite, the $NN$ interaction is simplified as contact potential, which generates the point-coupling RMF (PCRMF) models~\cite{nikolaus92}.

In the nonlinear RMF (NLRMF) model, the baryons interact with each other via exchanging various light mesons, including scalar-isoscalar meson ($\sigma$), vector-isoscalar meson ($\omega$), vector-isovector meson ($\rho$), and strange scalar and vector mesons ($\sigma^*$ and $\phi$)~\cite{ring96,vretenar05,meng06,dutra14,lu11,liu18}. The baryons considered in the present calculation are nucleons ($n$ and $p$) and hyperons ($\Lambda,~\Sigma,~\Xi$). The Lagrangian density of NLRMF model  is written as
\begin{equation}
	\begin{aligned}
		\mathcal{L}_{\rm NL}=& \sum_{B}\overline{\psi}_B\left\{i\gamma^{\mu}\partial_{\mu}-\left(M_B-g_{\sigma B}\sigma-g_{\sigma^* B}\sigma^*\right)\right.\\
		&\left.-\gamma^{\mu}\left(g_{\omega B}\omega_{\mu}+g_{\phi B}\phi_{\mu}
		+\frac{1}{2}g_{\rho B}\vec{\tau}\vec{\rho}_{\mu}\right)\right\}\psi_B\\
		&+\frac{1}{2}\partial^{\mu}\sigma\partial_{\mu}\sigma-\frac{1}{2}m_{\sigma}^2\sigma^2-\frac{1}{3}g_{2}\sigma^3
		-\frac{1}{4}g_{3}\sigma^4\\
		&+\frac{1}{2}\partial^{\mu}\sigma^*\partial_{\mu}\sigma^*-\frac{1}{2}m_{\sigma^*}^2\sigma^{*2}\\
		&-\frac{1}{4}W^{\mu\nu}W_{\mu\nu}+\frac{1}{2}m_{\omega}^2\omega^{\mu}\omega_{\mu}+\frac{1}{4}c_3\left(\omega^{\mu}\omega_{\mu}\right)^2\\
		&-\frac{1}{4}\Phi^{\mu\nu}\Phi_{\mu\nu}+\frac{1}{2}m_{\phi}^2\phi^{\mu}\phi_{\mu}\\
		&-\frac{1}{4}\vec{R}^{\mu\nu}\vec{R}_{\mu\nu}+\frac{1}{2}m_{\rho}^2\vec{\rho}^{\mu}\vec{\rho}_{\mu}\\
		&+\Lambda_{\rm v}\left(g_{\omega}^2\omega^{\mu}\omega_{\mu}\right)\left(g_{\rho}^2\vec{\rho}^{\mu}\vec{\rho}_{\mu}\right),
	\end{aligned}
\end{equation}
where $\psi_B$ represents the wave function of baryons. $\sigma,~\sigma^*,~\omega_{\mu},~\phi_{\mu}~\vec{\rho}_\mu$ denote the fields of $\sigma,~\sigma^*,~\omega,~\phi$, and $\rho$ mesons, respectively. $W_{\mu\nu}$,  $\Phi_{\mu\nu}$, and $\vec{R}_{\mu\nu}$ are the anti-symmetry tensor fields of $\omega$, $\phi$, and $\rho$ mesons. Here $g_{\omega B}$ denotes the coupling constant between $\omega$ meson and baryon, while $g_\omega$ for the coupling strength between $\omega$ meson and nucleon.  To solve the nuclear many-body system in the framework of the NLRMF model, the mean-field approximation should be adopted, in which various mesons are treated as classical fields,
\begin{equation}
\begin{aligned}
	&\sigma~\rightarrow\left\langle\sigma\right\rangle\equiv\sigma,
	~\sigma^*~\rightarrow\left\langle\sigma^*\right\rangle\equiv\sigma^*,\\
	&~\omega_{\mu}\rightarrow\left\langle\omega_{\mu}\right\rangle\equiv\omega,
	~\phi_{\mu}\rightarrow\left\langle\phi_{\mu}\right\rangle\equiv\phi,\\
	&~\vec{\rho}_{\mu}\rightarrow\left\langle\vec{\rho}_{\mu}\right\rangle\equiv\rho.
\end{aligned}
\end{equation}
The space components of the vector mesons are removed in the parity conservation system. In addition, the spatial derivatives of baryons and mesons are neglected in the infinite nuclear matter due to the transformation invariance. Finally, with the Euler-Lagrange equation, the equations of motion for baryons and mesons are obtained,
\begin{equation}\label{1.3}
	\begin{aligned}
		&\left[i\gamma^{\mu}\partial_{\mu}-M_B^*-\gamma^0\left(g_{\omega B}\omega+g_{\phi B}\phi+\frac{g_{\rho B}}{2}\rho\tau_3\right)\right]\psi_B=0,\\
		&m_{\sigma}^2\sigma+g_{2}\sigma^2+g_{3}\sigma^3=\sum_B g_{\sigma B}\rho^{s}_B,\\
		&m_{\sigma^*}^2\sigma^*=\sum_B g_{\sigma^* B}\rho^{s}_B,\\
		&m_{\omega}^2\omega+c_{3}\omega^3+2\Lambda_{\rm v}\left(g_{\omega }^2\omega\right)(g_{\rho}^2\rho^2)=\sum_B g_{\omega B}\rho^{v}_{B},\\
		&m_{\phi}^2\phi=\sum_B g_{\phi B}\rho^{v}_{B},\\
		&m_{\rho}^2\rho+2\Lambda_{\rm v}(g_{\omega }^2\omega^2)(g_{\rho }^2\rho)=\sum_B \frac{g_{\rho B}}{2}\rho^{v3}_{B},
	\end{aligned}
\end{equation}
where  $\tau_3$ is the isospin third component of the baryon species $B$. The scalar ($s$), vector densities ($v$), and their isospin components are defined as,
\begin{align}\label{1.4}
	\rho^s_B=\left\langle\overline{\psi}_B\psi_B\right\rangle,~~~~
	&\rho^{s3}_B=\left\langle\overline{\psi}_B\tau_3\psi_B\right\rangle,\nonumber\\
	\rho^v_{B}=\left\langle\psi^{\dag}_B\psi_B\right\rangle,~~~~
	&\rho^{v3}_B=\left\langle\psi^{\dag}_B\tau_3\psi_B\right\rangle.
\end{align}
The effective masses of baryons in Eq.~\eqref{1.3} are dependent on the scalar mesons $\sigma$ and $\sigma^*$,
\begin{align}
	M_B^{*}&=M_B-g_{\sigma B}\sigma-g_{\sigma^* B}\sigma^*.
\end{align}
The corresponding effective energies of baryons {take} the following form because of the mass-energy relation,
\begin{gather}\label{fere}
	E_{FB}^{*}=\sqrt{k_{FB}^2+(M_B^{*})^2},
\end{gather}
where $k_{FB}$ is the Fermi momentum of baryons. 

With the energy-momentum tensor in a uniform system, the energy density, $\mathcal{E}$ and pressure, $P$ of infinite nuclear matter are obtained respectively as~\cite{dutra14}
\begin{align}
	\mathcal{E}_{\rm NL}=&\frac{\gamma}{16\pi^2}\sum_{B}\left[k_{FB}E_{FB}^{*}\left(2k_{FB}^2
	+{M_B^{*}}^2\right)\right.\nonumber\\
	&\left.+{M_B^{*}}^4{\rm ln}\frac{M_B^{*}}{k_{FB}+E_{FB}^{*}}\right]\nonumber\\
	&+\frac{1}{2}m_{\sigma}^2\sigma^2+\frac{1}{3}g_{2}\sigma^3+\frac{1}{4}g_{3}\sigma^4+\frac{1}{2}m_{\sigma^*}^2{\sigma^*}^2\nonumber\\
	&+\frac{1}{2}m_{\omega}^2\omega^2+\frac{3}{4}c_{3}\omega^4+\frac{1}{2}m_{\rho}^2\rho^2+3\Lambda_{\rm v}\left(g_{\omega}^2\omega^{2}\right)\left(g_{\rho}^2\rho^{2}\right),\\ 
	P_{\rm NL}=&\frac{\gamma}{48\pi^2}\sum_{B}\left[3{M_B^{*}}^{4}{\rm ln}\frac{k_{FB}+E_{FB}^{*}}{M_B^{*}}\right.\nonumber\\
	&\left.+k_{FB}^2\left(2k_{FB}^2-3{M_B^{*}}^2\right)E_{FB}^{*}\right]\nonumber\\
	& -\frac{1}{2}m_{\sigma}^2\sigma^2
	-\frac{1}{3}g_{2}\sigma^3-\frac{1}{4}g_{3}\sigma^4-\frac{1}{2}m_{\sigma^*}^2{\sigma^*}^2\nonumber\\
	&+\frac{1}{2}m_{\omega}^2\omega^2+\frac{c_{3}}{4}\omega^4+\frac{1}{2}m_{\rho}^2\rho^2+\Lambda_{\rm v}\left(g_{\omega}^2\omega^{2}\right)\left(g_{\rho}^2\rho^{2}\right).
\end{align}
where $\gamma=2$ is the spin degeneracy factor. 

The outer core part of a neutron star, which almost dominates its mass and radius, is usually treated as the uniform matter consisting of  baryons and leptons.  Therefore their chemical potentials are very important, that are derived from the thermodynamics equations at zero temperature, 
\begin{equation}
	\begin{gathered}
		\mu_B=\sqrt{k_{FB}^2+M_B^{*2}}+g_{\omega B}\omega+g_{\phi B}\phi+\frac{g_{\rho B}}{2}\tau_3\rho,\\
		\mu_l=\sqrt{k_{Fl}^2+m_l^{2}},~~~~~l=e,~\mu.
	\end{gathered}
\end{equation} 

\subsection{The Density-dependent Relativistic Mean-field Model} 
In the DDRMF model, the Lagrangian density of nuclear many-body system has the similar form as that of NLRMF model,
	\begin{align}
		\mathcal{L}_{\rm DD}
		=&\sum_{B}\overline{\psi}_B\left[\gamma^{\mu}\left(i\partial_{\mu}-\Gamma_{\omega B}(\rho_B)\omega_{\mu}\right.\right.\nonumber\\
		&\left.\left.-\Gamma_{\phi B}(\rho_B)\phi_{\mu}-\frac{\Gamma_{\rho B}(\rho_B)}{2}\vec{\rho}_{\mu}\vec{\tau}\right)\right.\nonumber\\
		&\left.-\left(M_B-\Gamma_{\sigma B}(\rho_B)\sigma-\Gamma_{\sigma^* B}(\rho_B)\sigma^*-\Gamma_{\delta B}(\rho_B)\vec{\delta}\vec{\tau}\right)\right]\psi_B\nonumber\\
		&+\frac{1}{2}\left(\partial^{\mu}\sigma\partial_{\mu}\sigma-m_{\sigma}^2\sigma^2\right)
		+\frac{1}{2}\left(\partial^{\mu}\sigma^*\partial_{\mu}\sigma^*-m_{\sigma^*}^2{\sigma^*}^2\right)\nonumber\\
		&+\frac{1}{2}\left(\partial^{\mu}\vec{\delta}\partial_{\mu}\vec{\delta}-m_{\delta}^2\vec{\delta}^2\right)
		-\frac{1}{4}W^{\mu\nu}W_{\mu\nu}+\frac{1}{2}m_{\omega}^2\omega_{\mu}\omega^{\mu}\nonumber\\
		&-\frac{1}{4}\Phi^{\mu\nu}\Phi_{\mu\nu}+\frac{1}{2}m_{\phi}^2\phi_{\mu}\phi^{\mu}
		-\frac{1}{4}\vec{R}^{\mu\nu}\vec{R}_{\mu\nu}+\frac{1}{2}m_{\rho}^2\vec{\rho}_{\mu}\vec{\rho}^{\mu},
	\end{align}
where a scalar-isovector meson ($\delta$) is also introduced due to some parameterization. The coupling constants of $\sigma$ and $\omega$ mesons are usually expressed as a fraction function of the total vector density, $\rho_B=\sum_B\rho^v_B$. In most of DDRMF parametrizations, such as DD2~\cite{dd2}, DD-ME1~\cite{ddme1}, DD-ME2~\cite{ddme2}, DDME-X~\cite{ddmex}, PKDD~\cite{pkdd}, TW99~\cite{tw99}, and DDV, DDVT, DDVTD~\cite{ddv} , they are assumed as,
\begin{equation}\label{cdd}
	\Gamma_{iN}(\rho_B)=\Gamma_{iN}(\rho_{B0})f_i(x)
\end{equation}
with
\begin{equation}\label{cdd}
f_i(x)=a_i\frac{1+b_i(x+d_i)^2}{1+c_i(x+d_i)^2},~x=\rho_B/\rho_{B0},
\end{equation}
for $i=\sigma,~\omega$. $\rho_{B0}$ is the saturation density of symmetric nuclear matter. Five constraints on the coupling constants $f_i(1)=1,~f_i^{''}(0)=0,~f_{\sigma}^{''}(1)=f_{\omega}^{''}(1)$ can reduce the numbers of independent parameters to three in Eq.~\eqref{cdd}. The first two constraints lead to
\begin{gather}
	a_i=\frac{1+c_i(1+d_i)^2}{1+b_i(1+d_i)^2},~~~~3c_id_i^2=1.
\end{gather}
For the isovector mesons $\rho$ and $\delta$, their density-dependent coupling constants are assumed to be,
\begin{gather}
	\Gamma_{iN}(\rho_B)=\Gamma_{iN}(\rho_{B0}){\rm exp}[-a_i(x-1)].
\end{gather}
While in other parametrizations, such as DD-LZ1~\cite{ddlz1}, the coefficient in front of fraction function, $\Gamma_i$ is fixed at $\rho_B=0$ for $i=\sigma,~\omega$:
\begin{gather}\label{1.5}
	\Gamma_{iN}(\rho_B)=\Gamma_{iN}(0)f_i(x).
\end{gather}
There are only four constraint conditions as $f_i(0)=1$ and $f''_i(0)=0$ for $\sigma$ and $\omega$ coupling constants in DD-LZ1 set. The constraint $f''_{\sigma}(1)=f''_{\omega}(1)$ in previous parameter sets was removed in DD-LZ1 parametrization. For $\rho$ meson, its coupling constant is also changed accordingly as 
\begin{gather}\label{1.6}
	\Gamma_{\rho N}(\rho_B)=\Gamma_{\rho N}(0){\rm exp}(-a_{\rho}x).
\end{gather}
Within the mean-field approximation, the meson fields are treated as the classical fields,
$
\left\langle\sigma\right\rangle=\sigma,~\left\langle\sigma^*\right\rangle=\sigma^*,~\left\langle\omega_{\mu}\right\rangle=\omega,~\left\langle\phi_{\mu}\right\rangle=\phi,
~\left\langle\vec{\rho}_{\mu}\right\rangle=\rho,~\langle\delta\rangle=\delta.
$
Together with the Euler-Lagrange equations, the equations of motion for baryons and mesons are given by :
\begin{align}
	\label{1.8}
	&\left[i\gamma^{\mu}\partial_{\mu}-\gamma^0\left(\Gamma_{\omega B}(\rho_B)\omega+\Gamma_{\phi B}(\rho_B)\phi\right.\right.\nonumber\\
	&\left.\left.+\frac{\Gamma_{\rho B}(\rho_B)}{2}\rho\tau_3+\Sigma_R(\rho_B)\right)-M_B^{*}\right]\psi_B=0.\nonumber\\
	&m_{\sigma}^2\sigma=\sum_B\Gamma_{\sigma B}(\rho_B)\rho^s_B,\nonumber\\
	&m_{\sigma^*}^2\sigma^*=\sum_B\Gamma_{\sigma^* B}(\rho_B)\rho^s_B,\nonumber\\
	&m_{\omega}^2\omega=\sum_B\Gamma_{\omega B}(\rho_B)\rho^v_B,\nonumber\\
	&m_{\phi}^2\phi=\sum_B\Gamma_{\phi B}(\rho_B)\rho^v_B,\nonumber\\
	&m_{\rho}^2\rho=\sum_B\frac{\Gamma_{\rho B}(\rho_B)}{2}\rho^{v3}_B,\nonumber\\
	&m_{\delta}^2\delta=\sum_B\Gamma_{\delta B}(\rho_B)\rho^{s3}_B.
\end{align}
Comparing to the NLRMF model, an additional term about the rearrangement contribution, $\Sigma_R$ will be introduced into the vector potential in Eq.~\eqref{1.8} due to the vector density-dependence of coupling constants,
	\begin{align}
		\Sigma_R(\rho_B)=&-\frac{\partial\Gamma_{\sigma B}(\rho_B)}{\partial\rho_B}\sigma\rho^s_B
		-\frac{\partial\Gamma_{\sigma^* B}(\rho_B)}{\partial\rho_B}\sigma^*\rho^s_B\nonumber\\
		&-\frac{\partial\Gamma_{\delta B}(\rho_B)}{\partial\rho_B}\delta\rho^{s3}_B
		+\frac{1}{2}\frac{\partial\Gamma_{\rho B}(\rho_B)}{\partial\rho_B}\rho\rho^{v3}_B\nonumber\\
		&+\left[\frac{\partial\Gamma_{\omega B}(\rho_B)}{\partial\rho_B}\omega+\frac{\partial\Gamma_{\phi B}(\rho_B)}{\partial\rho_B}\phi\right]\rho^v_B,
	\end{align}
where the scalar, vector densities, and their isospin components {take} the same forms as shown in Eq.~\eqref{1.4}.
The effective masses of baryons in Eq.~\eqref{1.8} are dependent on the scalar mesons $\sigma$, $\sigma^*$ and $\delta$,
\begin{align}
	M_B^{*}&=M_B-\Gamma_{\sigma B}(\rho_B)\sigma-\Gamma_{\sigma^* B}(\rho_B)\sigma^*-\Gamma_{\delta B}(\rho_B)\delta\tau_3
\end{align}
and the corresponding effective energies of baryons {take}  the same form as Eq. \eqref{fere}.

The energy density, $\mathcal{E}$ and pressure, $P$ of nuclear matter in DDRMF model are obtained respectively as
\begin{equation}
	\begin{aligned}
		\mathcal{E}_{\rm DD}=&\frac{1}{2}m_{\sigma}^2\sigma^2+\frac{1}{2}m_{\sigma^*}^2{\sigma^*}^2
		-\frac{1}{2}m_{\omega}^2\omega^2-\frac{1}{2}m_{\phi}^2\phi^2\\
		&-\frac{1}{2}m_{\rho}^2\rho^2+\frac{1}{2}m_{\delta}^2\delta^2+\Gamma_{\omega B}(\rho_B)\omega\rho_B\\
		&+\Gamma_{\phi B}(\rho_B)\phi\rho_B +\frac{\Gamma_{\rho}(\rho_B)}{2}\rho\rho_3+\mathcal{E}_{\rm kin}^B,\\
		P_{\rm DD}=
		&\rho_B\Sigma_{R}(\rho_B)-\frac{1}{2}m_{\sigma}^2\sigma^2-\frac{1}{2}m_{\sigma^*}^2{\sigma^*}^2\\
		&+\frac{1}{2}m_{\omega}^2\omega^2+\frac{1}{2}m_{\phi}^2\phi^2+\frac{1}{2}m_{\rho}^2\rho^2\\
		&-\frac{1}{2}m_{\delta}^2\delta^2
		+P_{\rm kin}^B,
	\end{aligned}
\end{equation}
where, the contributions from kinetic energy are 
\begin{align}
	\mathcal{E}_{\rm kin}^B&=\frac{\gamma}{2\pi^2}\int_{0}^{k_{FB}}k^2\sqrt{k^2+{M_B^{*}}^{2}}dk\nonumber\\
	&=\frac{\gamma}{16\pi^2}\left[k_{FBi}E_{FB}^{*}\left(2k_{FB}^2+{M_B^{*}}^2\right)\right.\nonumber\\
	&\left.+{M_B^{*}}^4{\rm ln}\frac{M_B^{*}}{k_{FB}+E_{FB}^{*}}\right], \\
	P_{\rm kin}^B&=\frac{\gamma}{6\pi^2}\int_{0}^{k_{FB}}\frac{k^4 dk}{\sqrt{k^2+{M_B^{*}}^{2}}}\nonumber\\
	&=\frac{\gamma}{48\pi^2}\left[k_{FB}\left(2k_{FB}^2-3{M_B^{*}}^2\right)E_{FB}^{*}\right.\nonumber\\
	&\left.+3{M_B^{*}}^{4}{\rm ln}\frac{k_{FB}+E_{FB}^{*}}{M_B^{*}}\right].
\end{align}
$\gamma=2$ is the spin degeneracy factor. 

The chemical potentials of baryons and leptons are 
\begin{align}\label{2.1}
	\mu_{B}&=\sqrt{k_{FB}^2+M_B^{*2}}+\left[\Gamma_{\omega B}(\rho_B)\omega+\Gamma_{\phi B}(\rho_B)\phi\right.\nonumber\\
	&\left.+\frac{\Gamma_{\rho B}(\rho_B)}{2}\rho\tau_3+\Sigma_R(\rho_B)\right],\\\nonumber
	\label{2.2}
	\mu_l&=\sqrt{k_{Fl}^2+m_l^{2}}.
\end{align}

\subsection{The formulas about neutron star} 
In the uniform hyperonic star matter, the compositions of baryons and leptons are determined by the requirements of charge neutrality and $\beta$-equilibrium conditions. All baryon octets $(n,~p,~\Lambda,~\Sigma^-,~\Sigma^0,~\Sigma^+,~\Xi^-,~\Xi^0)$ and leptons ($e,~\mu$) will be included in this work. The $\beta$-equilibrium conditions can be expressed by~\cite{glendenning85,shen02}
\begin{equation}\label{1.23}
	\begin{gathered} 
		\mu_p=\mu_{\Sigma^+}=\mu_n-\mu_e,\\
		\mu_{\Lambda}=\mu_{\Sigma^0}=\mu_{\Xi^0}=\mu_n,\\
		\mu_{\Sigma^-}=\mu_{\Xi^-}=\mu_n+\mu_e,\\
		\mu_{\mu}=\mu_e.
	\end{gathered}
\end{equation}
The charge neutrality condition has the following form,
\begin{gather}\label{1.24}
	\rho^v_{p}+\rho^v_{\Sigma^+}=\rho^v_e+\rho^v_{\mu}+\rho^v_{\Sigma^-}+\rho^v_{\Xi^-}.
\end{gather}

The total energy density and pressure of neutron star matter will be obtained as a function of baryon density within the constraints of Eqs.~\eqref{1.23} and \eqref{1.24}. The Tolman-Oppenheimer-Volkoff (TOV) equation describes a spherically symmetric star in the gravitational equilibrium from general relativity~\cite{oppenheimer39,tolman39},
\begin{gather}\label{tov}
	\frac{dP}{dr}=-\frac{GM(r)\mathcal{E}(r)}{r^2}\frac{\left[1+\frac{P(r)}{\mathcal{E}(r)}\right]\left[1+\frac{4\pi r^3P(r)}{M(r)}\right]}{1-\frac{2GM(r)}{r}},\\\nonumber
	\frac{dM(r)}{dr}=4\pi r^2\mathcal{E}(r),
\end{gather}
where $P$ and $M$ are the pressure and mass of a neutron star at the position $r$. Furthermore, the tidal deformability becomes a typical property of a neutron star after the observation of the gravitational wave from a binary neutron-star (BNS) merger, which characterizes the deformation of a compact object in an external gravity field generated by another star. The tidal deformability of a neutron star {is} reduced as a dimensionless form~\cite{hinderer08,postnikov10},
\begin{gather}\label{dt}
	\Lambda=\frac{2}{3}k_2C^{-5}.
\end{gather}
where $C=GM/R$ is the compactness parameter. The second order Love number $k_2$ is given by
\begin{align}\label{lneq}
	k_2=&\frac{8C^5}{5}(1-2C)^2\left[2+2C(\mathcal{Y}_R-1)-\mathcal{Y}_R\right]\nonumber\\
	&\Big\{2C\left[6-3\mathcal{Y}_R+3C(5\mathcal{Y}_R-8)\right]\nonumber\\
	&+4C^3\left[13-11\mathcal{Y}_R+C(3\mathcal{Y}_R-2)+2C^2(1+\mathcal{Y}_R)\right]\nonumber\\
	&+3(1-2C)^2\left[2-\mathcal{Y}_R+2C(\mathcal{Y}_R-1){\rm ln}(1-2C)\right]\Big\}^{-1}.
\end{align}
Here, $\mathcal{Y}_R=y(R)$. $y(r)$ satisfies the following first-order differential equation,
\begin{equation}
	r\frac{d y(r)}{dr} + y^2(r)+y(r)F(r) + r^2Q(r)=0,
\end{equation}
where $F(r)$ and $Q(r)$ are functions related to the pressure and energy density
\begin{align}
	F(r) & = \left[1-\frac{2M(r)}{r}\right]^{-1} 
	\left\{1-4\pi r^2[\mathcal{E}(r)-P(r)]\right\} ,\\ 
	\nonumber 
	r^2Q(r) & =  \left\{4\pi r^2 \left[5\mathcal{E}(r)+9P(r)+\frac{\mathcal{E}(r)
		+P(r)}{\frac{\partial P}{\partial \mathcal{E}}(r)}\right]-6\right\}\\ 
	\nonumber 
	&\times \left[1-\frac{2M(r)}{r}\right]^{-1} -\left[\frac{2M(r)}{r} +2\times4\pi r^2 P(r) \right]^2\\ 
	\nonumber 
	 &\times \left[1-\frac{2M(r)}{r}\right]^{-2} .
\end{align}
The second Love number corresponds to the initial condition $y(0)=2$. It is also related to the speed of sound in compact matter, $c_s$,
\begin{gather}
	c_s^2=\frac{\partial P(\varepsilon)}{\partial{\mathcal{E}}}.
\end{gather}

\section{RESULTS AND DISCUSSIONS}
\subsection{The RMF Parametrizations}
In the present work, four popular parameter sets, TM1~\cite{tm1}, NL3~\cite{nl3}, IUFSU~\cite{iufsu}, and BigApple~\cite{bigapple} in the NLRMF model are used to describe the uniform matter, which was generated by fitting the ground state properties of several stable nuclei. The nonlinear term of the vector meson was considered in the TM1 set to reproduce the nucleon vector potential from the RBHF model. The NL3 set can generate a large maximum mass of a neutron star, while its radius is large~\cite{ju21b}. The $\omega$-$\rho$ meson coupling term was included in the IUFSU set. The BigApple set can lead to a large maximum mass of a neutron star with a small radius due to the smaller slope of the symmetry energy~\cite{bigapple}. The masses of nucleons and mesons, and the coupling constants between nucleon and mesons in NLRMF models, TM1 \cite{tm1}, IUFSU \cite{iufsu}, BigApple \cite{bigapple}, NL3 \cite{nl3} are tabulated in Table~\ref{table.1}.  

\begin{table}[htb]
	\small
	\centering
	\caption{ Masses of nucleons and mesons, meson coupling constants in various NLRMF models}\label{table.1}
	\scalebox{0.85}{
		\begin{tabular}{lccccc}
			\hline\hline
			&                     ~~~&NL3~\cite{nl3}       ~~~&BigApple~\cite{bigapple}    ~~~&TM1~\cite{tm1}        ~~~&IUFSU~\cite{iufsu}     \\
			\hline                          
			&$m_n[\rm MeV]$           ~~~&939.000   ~~~&939.000     ~~~&938.000    ~~~&939.000    \\
			&$m_p[\rm MeV]$           ~~~&939.000   ~~~&939.000     ~~~&938.000    ~~~&939.000    \\
			&$m_{\sigma}[\rm MeV]$    ~~~&508.194   ~~~&492.730     ~~~&511.198    ~~~&491.500    \\
			&$m_{\omega}[\rm MeV]$    ~~~&782.501   ~~~&782.500     ~~~&783.000    ~~~&782.500    \\
			&$m_{\rho}[\rm MeV]$      ~~~&763.000   ~~~&763.000     ~~~&770.000    ~~~&763.000    \\
			&$g_{\sigma}$             ~~~&10.217    ~~~&9.6699      ~~~&10.0289    ~~~&9.9713     \\
			&$g_{\omega}$             ~~~&12.868    ~~~&12.316      ~~~&12.6139    ~~~&13.0321    \\
			&$g_{\rho}$               ~~~&8.948     ~~~&14.1618     ~~~&9.2644     ~~~&13.5899    \\
			&$g_2(\rm fm^{-1})$       ~~~&10.431    ~~~&11.9214     ~~~&7.2325     ~~~&8.4929     \\
			&$g_3$                    ~~~&-28.885   ~~~&-31.6796    ~~~&0.6183     ~~~&0.4877     \\
			&$c_3$                    ~~~&0.000     ~~~&2.6843      ~~~&71.3075    ~~~&144.2195   \\
			&$\Lambda_V$              ~~~&0.000     ~~~&0.047471    ~~~&0.000      ~~~&0.046     \\
			\hline\hline
	\end{tabular}}
\end{table}

For the DDRMF models, the parametrizations,  DD2~\cite{dd2}, DD-ME1~\cite{ddme1}, DD-ME2~\cite{ddme2}, DDME-X~\cite{ddme2}, PKDD~\cite{pkdd}, TW99~\cite{tw99}, DDV, DDVT, DDVTD~\cite{ddv}, and DD-LZ1~\cite{ddlz1} are listed in Table~\ref{table.2}. TW99, DD2, DD-ME1, DD-ME2, and PKDD parameter sets were obtained by reproducing the ground-state properties of different nuclei before 2005. In recent years, the DDME-X, DD-LZ1,  DDV, DDVT, and DDVTD sets were brought out with various considerations. In DDVT and DDVTD sets, the tensor coupling between the vector meson and nucleon was included. The scalar-isovector meson, $\delta$ was taken into account in DDVTD. The coefficients of meson coupling constants, $\Gamma_i$ in DD-LZ1 are the values at zero density, while other parameter sets are dependent on the values at nuclear saturation densities. 

\begin{table*}[htb]
	\footnotesize
	\centering
	\caption{ Masses of nucleons and mesons, meson coupling constants, and the nuclear saturation densities in various DDRMF models.}\label{table.2}
	\scalebox{0.65}{
		\begin{tabular}{lcccccccccccccc}
			\hline\hline
			&DD-LZ1~\cite{ddlz1}     ~~&                               ~~&DD-MEX~\cite{ddmex}      ~~&DD-ME2~\cite{ddme2}    ~~&DD-ME1~\cite{ddme1}    ~~&DD2~\cite{dd2}         ~~&PKDD~\cite{pkdd}        ~~&TW99~\cite{tw99}        ~~&DDV~\cite{ddv}         ~~&DDVT~\cite{ddv}       ~~&DDVTD~\cite{ddv}          \\
			\hline
			$m_n[\rm MeV]$           &938.900000 ~~&$m_n$                          ~~&939.0000   ~~&939.0000  ~~&939.0000   ~~&939.56536    ~~&939.5731    ~~&939.0000    ~~&939.565413  ~~&939.565413  ~~&939.565413     \\
			$m_p[\rm MeV]$           &938.900000 ~~&$m_p$                          ~~&939.0000   ~~&939.0000  ~~&939.0000   ~~&938.27203    ~~&938.2796    ~~&939.0000    ~~&938.272081  ~~&938.272081  ~~&938.272081     \\
			$m_{\sigma}[\rm MeV]$    &538.619216 ~~&$m_{\sigma}$                   ~~&547.3327   ~~&550.1238  ~~&549.5255   ~~&546.212459   ~~&555.5112    ~~&550.0000    ~~&537.600098  ~~&502.598602  ~~&502.619843     \\
			$m_{\omega}[\rm MeV]$    &783.0000   ~~&$m_{\omega}$                   ~~&783.0000   ~~&783.0000  ~~&783.0000   ~~&783.0000     ~~&783.0000    ~~&783.0000    ~~&783.0000    ~~&783.0000    ~~&783.0000       \\
			$m_{\rho}[\rm MeV]$      &769.0000   ~~&$m_{\rho}$                     ~~&763.0000   ~~&763.0000  ~~&763.0000   ~~&763.0000     ~~&763.0000    ~~&763.0000    ~~&763.0000    ~~&763.0000    ~~&763.0000       \\
			$m_{\delta}[\rm MeV]$    &---        ~~&$m_{\delta}$                   ~~&---        ~~&---       ~~&---        ~~&---          ~~&---         ~~&---         ~~&---         ~~&---         ~~&980.0000       \\
			$\Gamma_{\sigma}(0)$     &12.001429  ~~&$\Gamma_{\sigma}(\rho_{B0})$   ~~&10.7067    ~~&10.5396   ~~&10.4434    ~~&10.686681    ~~&10.7385     ~~&10.72854    ~~&10.136960   ~~&8.382863    ~~&8.379269       \\
			$\Gamma_{\omega}(0)$     &14.292525  ~~&$\Gamma_{\omega}(\rho_{B0})$   ~~&13.3388    ~~&13.0189   ~~&12.8939    ~~&13.342362    ~~&13.1476     ~~&13.29015    ~~&12.770450   ~~&10.987106   ~~&10.980433      \\
			$\Gamma_{\rho}(0)$       &15.150934  ~~&$\Gamma_{\rho}(\rho_{B0})$     ~~&7.2380     ~~&7.3672    ~~&7.6106     ~~&7.25388      ~~&8.5996      ~~&7.32196     ~~&7.84833     ~~&7.697112    ~~&8.06038        \\
			$\Gamma_{\delta}(0)$     &---        ~~&$\Gamma_{\delta}(\rho_{B0})$   ~~&---        ~~&---       ~~&--         ~~&---          ~~&---         ~~&---         ~~&---         ~~&---         ~~&0.8487420      \\
			\hline                                                                            
			$\rho_{B0}[\rm fm^{-3}]$ &0.1581     ~~&$\rho_{B0}$    ~~&0.1520     ~~&0.1520    ~~&0.1520     ~~&0.1490       ~~&0.1496      ~~&0.1530      ~~&0.1511      ~~&0.1536      ~~&0.1536         \\
			\hline                                                                                    
			$a_{\sigma}$             &1.062748   ~~&$a_{\sigma}$                   ~~&1.3970     ~~&1.3881    ~~&1.3854     ~~&1.357630     ~~&1.327423    ~~&1.365469    ~~&1.20993     ~~&1.20397     ~~&1.19643        \\
			$b_{\sigma}$             &1.763627   ~~&$b_{\sigma}$                   ~~&1.3350     ~~&1.0943    ~~&0.9781     ~~&0.634442     ~~&0.435126    ~~&0.226061    ~~&0.21286844  ~~&0.19210314  ~~&0.19171263     \\
			$c_{\sigma}$             &2.308928   ~~&$c_{\sigma}$                   ~~&2.0671     ~~&1.7057    ~~&1.5342     ~~&1.005358     ~~&0.691666    ~~&0.409704    ~~&0.30798197  ~~&0.27773566  ~~&0.27376859     \\
			$d_{\sigma}$             &0.379957   ~~&$d_{\sigma}$                   ~~&0.4016     ~~&0.4421    ~~&0.4661     ~~&0.575810     ~~&0.694210    ~~&0.901995    ~~&1.04034342  ~~&1.09552817  ~~&1.10343705     \\
			$a_{\omega}$             &1.059181   ~~&$a_{\omega}$                   ~~&1.3936     ~~&1.3892    ~~&1.3879     ~~&1.369718     ~~&1.342170    ~~&1.402488    ~~&1.23746     ~~&1.16084     ~~&1.16693        \\
			$b_{\omega}$             &0.418273   ~~&$b_{\omega}$                   ~~&1.0191     ~~&0.9240    ~~&0.8525     ~~&0.496475     ~~&0.371167    ~~&0.172577    ~~&0.03911422  ~~&0.04459850  ~~&0.02640016     \\
			$c_{\omega}$             &0.538663   ~~&$c_{\omega}$                   ~~&1.6060     ~~&1.4620    ~~&1.3566     ~~&0.817753     ~~&0.611397    ~~&0.344293    ~~&0.07239939  ~~&0.06721759  ~~&0.04233010     \\
			$d_{\omega}$             &0.786649   ~~&$d_{\omega}$                   ~~&0.4556     ~~&0.4775    ~~&0.4957     ~~&0.638452     ~~&0.738376    ~~&0.983955    ~~&2.14571442  ~~&2.22688558  ~~&2.80617483     \\
			$a_{\rho}$               &0.776095   ~~&$a_{\rho}$                     ~~&0.6202     ~~&0.5647    ~~&0.5008     ~~&0.518903     ~~&0.183305    ~~&0.5150      ~~&0.35265899  ~~&0.54870200  ~~&0.55795902     \\
			$a_{\delta}$             &---        ~~&$a_{\delta}$                   ~~&---        ~~&---       ~~&---        ~~&---          ~~&---         ~~&---         ~~&---         ~~&---         ~~&0.55795902     \\
			\hline\hline
	\end{tabular}}
\end{table*}

The saturation properties of symmetric nuclear matter calculated with different RMF effective interactions are listed in Table~\ref{table.3}, i.e. the saturation density, $\rho_0$, the binding energy per nucleon, $E/A$, incompressibility, $K_0$, symmetry energy, $E_\text{sym}$, the slope of symmetry energy, $L$, and the effective neutron and proton masses, $M^*_n$ and $M^*_p$.  The saturation densities of nuclear matter are in the region of $0.145-0.158$ fm$^{-3}$. The binding energies per nucleon at saturation density are around $-16.5$ MeV. The incompressibilities of nuclear matter are $230\sim280$ MeV. The symmetry energies are $31\sim37$ MeV. The slopes of the symmetry energy from these RMF interactions have large uncertainties, from $40-120$ MeV, since there are still not too many experimental constraints about the neutron skin~\cite{li05,li08,tsang09,danielewics14,tsang19,li19}. The effective mass differences between neutron and proton in this table are caused by the differences of their free masses.

\begin{table*}[htb]
	\small
	\centering
	\caption{Nuclear matter properties at saturation density generated by NLRMF and DDRMF parameterizations.}\label{table.3}
	\scalebox{0.85}{
		\begin{tabular}{lcccccccccccccccc}
			\hline\hline
			&$\rho_{B0}[\rm fm^{-3}]$ ~~&$E/A[\rm MeV]$   ~~&$K_0[\rm MeV]$  ~~&$E_{\rm sym}[\rm MeV]$  ~~&$L_0[\rm MeV]$ ~~&$M_n^{*}/M$ ~~&$M_p^{*}/M$\\
			\hline
			NL3          &0.1480  ~~&-16.2403  ~~&269.9605   ~~&37.3449   ~~&118.3225   ~~&0.5956   ~~&0.5956 \\
			BigApple     &0.1545  ~~&-16.3436  ~~&226.2862   ~~&31.3039   ~~&39.7407    ~~&0.6103   ~~&0.6103 \\
			TM1          &0.1450  ~~&-16.2631  ~~&279.5858   ~~&36.8357   ~~&110.6082   ~~&0.6348   ~~&0.6348 \\
			IUFSU        &0.1545  ~~&-16.3973  ~~&230.7491   ~~&31.2851   ~~&47.1651    ~~&0.6095   ~~&0.6095 \\
			\hline\hline
			DD-LZ1       &0.1581  ~~&-16.0598  ~~&231.1030  ~~&31.3806  ~~&42.4660  ~~&0.5581  ~~&0.5581  \\
			DD-MEX       &0.1519  ~~&-16.0973  ~~&267.3819  ~~&32.2238  ~~&46.6998  ~~&0.5554  ~~&0.5554  \\
			DD-ME2       &0.1520  ~~&-16.1418  ~~&251.3062  ~~&32.3094  ~~&51.2653  ~~&0.5718  ~~&0.5718  \\
			DD-ME1       &0.1522  ~~&-16.2328  ~~&245.6657  ~~&33.0899  ~~&55.4634  ~~&0.5776  ~~&0.5776  \\
			DD2          &0.1491  ~~&-16.6679  ~~&242.8509  ~~&31.6504  ~~&54.9529  ~~&0.5627  ~~&0.5614  \\
			PKDD         &0.1495  ~~&-16.9145  ~~&261.7912  ~~&36.7605  ~~&90.1204  ~~&0.5713  ~~&0.5699  \\
			TW99         &0.1530  ~~&-16.2472  ~~&240.2022  ~~&32.7651  ~~&55.3095  ~~&0.5549  ~~&0.5549  \\
			DDV          &0.1511  ~~&-16.9279  ~~&239.9522  ~~&33.5969  ~~&69.6813  ~~&0.5869  ~~&0.5852  \\
			DDVT         &0.1536  ~~&-16.9155  ~~&239.9989  ~~&31.5585  ~~&42.3414  ~~&0.6670  ~~&0.6657  \\
			DDVTD        &0.1536  ~~&-16.9165  ~~&239.9137  ~~&31.8168  ~~&42.5829  ~~&0.6673  ~~&0.6660  \\
			\hline\hline
	\end{tabular}}
\end{table*}

For the hyperonic star matter with strangeness degree of freedom, the hyperon masses are chosen as $m_{\Lambda}=1115.68$ MeV, $m_{\Sigma^+}=1189.37$ MeV, $m_{\Sigma^0}=1192.64$ MeV, $m_{\Sigma^-}=1197.45$ MeV, $m_{\Xi^0}=1314.86$ MeV, and  $m_{\Xi^-}=1321.71$ MeV~\cite{zyla20}, while the masses of strange mesons, $\phi$ and $\sigma^*$ are taken as $m_{\phi}=1020$ MeV and $m_{\phi}=980$ MeV. We adopt the values from the SU(6) symmetry for the coupling constants between hyperons and vector mesons~\cite{dover84},
\begin{equation}
	\begin{gathered}
		\Gamma_{\omega\Lambda}=\Gamma_{\omega\Sigma}=2\Gamma_{\omega\Xi}=\frac{2}{3}\Gamma_{\omega N},\\
		2\Gamma_{\phi\Sigma}=\Gamma_{\phi\Xi}=-\frac{2\sqrt{2}}{3}\Gamma_{\omega N},~\Gamma_{\phi N}=0,\\
		\Gamma_{\rho\Lambda}=0,~\Gamma_{\rho\Sigma}=2\Gamma_{\rho\Xi}=2\Gamma_{\rho N},\\
		\Gamma_{\delta\Lambda}=0,~\Gamma_{\delta\Sigma}=2\Gamma_{\delta\Xi}=2\Gamma_{\delta N}.\\
	\end{gathered}  
\end{equation}
The coupling constants of hyperon and scalar mesons are constrained by the hyperon-nucleon potentials in symmetric nuclear matter, $U_Y^N$, which are defined by
\begin{equation}\label{pot}
	U_Y^N(\rho_{B0})=-R_{\sigma Y}\Gamma_{\sigma N}(\rho_{B0})\sigma_0+R_{\omega Y}\Gamma_{\omega N}(\rho_{B0})\omega_0,
\end{equation}
where $\Gamma_{\sigma N}(\rho_{B0}),~\Gamma_{\omega N}(\rho_{B0}),~\sigma_0,~\omega_0$ are the values of coupling strengths and $\sigma,~\omega$ meson fields at the saturation density. $R_{\sigma Y}$ and $R_{\omega Y}$ are defined as $R_{\sigma Y}=\Gamma_{\sigma Y}/\Gamma_{\sigma N}$ and $R_{\omega Y}=\Gamma_{\omega Y}/\Gamma_{\omega N}$. We choose the hyperon-nucleon potentials of $\Lambda,~\Sigma$ and $\Xi$ as $U_{\Lambda}^N=-30$ MeV, $U_{\Sigma}^N=+30$ MeV and $U_{\Xi}^N=-14$ MeV, respectively from the recent hypernuclei experimental observables~\cite{fortin17,hu22,hy_potential1}. 

The coupling constants between $\Lambda$ and $\sigma^*$ is generated by the value of the $\Lambda\Lambda$ potential in pure $\Lambda$ matter, $U_{\Lambda}^{\Lambda}$ at nuclear saturation density, which is given as
\begin{equation}
		\begin{aligned}
	U_{\Lambda}^{\Lambda}(\rho_{B0})=&-R_{\sigma\Lambda}\Gamma_{\sigma N}(\rho_{B0})\sigma_0-R_{\sigma^*\Lambda}\Gamma_{\sigma N}(\rho_{B0}){\sigma^*_0}\\
	&+R_{\omega Y}\Gamma_{\omega N}(\rho_{B0})\omega_0+R_{\phi\Lambda}\Gamma_{\omega N}(\rho_{B0})\phi_0,
	\end{aligned}
\end{equation}
We similarly define that $R_{\sigma^*\Lambda}=\Gamma_{\sigma^*\Lambda}/\Gamma_{\sigma N}$ and $R_{\phi\Lambda}=\Gamma_{\phi\Lambda}/\Gamma_{\omega N}$. $R_{\sigma^*\Lambda}$ is obtained from the $\Lambda-\Lambda$ potential as $U_{\Lambda}^{\Lambda}(\rho_{B0})=-10$ MeV, which was extracted from the $\Lambda$ bonding energies of double-$\Lambda$ hypernuclei. $R_{\phi\Lambda}=-\sqrt{2}/2$ is corresponding to the SU(6) symmetry broken case~\cite{fortin17}. Here, the coupling between the $\Sigma$, $\Xi$ hyperons and $\sigma^*$ mesons are set as $R_{\sigma^*\Xi}=0,~R_{\sigma^*\Sigma}=0$, since the information about their interaction is absent until now. The values of $R_{\sigma Y}$ and $R_{\sigma^*\Lambda}$ with above constraints in different RMF effective interactions are shown in Table~\ref{table.4}.

\begin{table}[htb]
	\small
	\centering
	\caption{The Coupling constants between hyperons and $\sigma$ meson, $g_{\sigma Y}$ and $\Lambda$-$\sigma^*$, $g_{\sigma^*\Lambda}$ in different RMF effective interactions.}\label{table.4}
	\scalebox{0.85}{
		\begin{tabular}{lcccccccccccc}
			\hline\hline
			&$R_{\sigma\Lambda}$ ~~&$R_{\sigma\Sigma}$  ~~&$R_{\sigma\Xi}$  ~~&$R_{\sigma^*\Lambda}$\\
			\hline
			NL3          &0.618896 ~~&0.460889  ~~&0.306814  ~~&0.84695\\
			BigApple     &0.616322 ~~&0.452837  ~~&0.305436  ~~&0.86313\\
			TM1          &0.621052 ~~&0.445880  ~~&0.307606  ~~&0.83710\\
			IUFSU        &0.616218 ~~&0.453006  ~~&0.305389  ~~&0.88802\\
			\hline\hline
			DD-LZ1       &0.610426 ~~&0.465708  ~~&0.302801  ~~&0.87595\\
			DD-MEX       &0.612811 ~~&0.469159  ~~&0.304011  ~~&0.86230\\
			DD-ME2       &0.609941 ~~&0.460706  ~~&0.302483  ~~&0.85758\\
			DD-ME1       &0.608602 ~~&0.457163  ~~&0.301777  ~~&0.85828\\
			DD2          &0.612743 ~~&0.466628  ~~&0.303937  ~~&0.86420\\
			PKDD         &0.610412 ~~&0.461807  ~~&0.302729  ~~&0.84965\\
			TW99         &0.612049 ~~&0.468796  ~~&0.303632  ~~&0.85818\\ 
			DDV          &0.607355 ~~&0.452777  ~~&0.301101  ~~&0.87979\\
			DDVT         &0.591179 ~~&0.399269  ~~&0.292391  ~~&0.92256\\
			DDVTD        &0.591108 ~~&0.399023  ~~&0.292352  ~~&0.92246\\
			\hline\hline
	\end{tabular}}
\end{table}

\subsection{Neutron Star and Hyperonic Star}\label{nsm}
{With the conditions of $\beta$ equilibrium and charge neutrality, the equations of state (EoSs) of neutron star matter including the neutron, proton, and leptons, i.e. the $P-\epsilon$ function in the NLRMF and  DDRMF models \textcolor{blue}{is} obtained in the panel (a) and panel (b) of Fig.~\eqref{fig.1}, which shows the pressures of neutron star matter as a function of energy density. For the inner crust part of a neutron star, the EoS of the non-uniform matter generated by TM1 parametrization with self-consistent Thomas-Fermi approximation is adopted~\cite{bao15}. In the core region of a neutron star, the EoSs of the uniform matter are calculated with various NLRMF and DDRMF parameter sets. {Furthermore,  the joint constraints on the EoS extracted from the GW170817 and GW190814 are shown as a shaded band~\cite{abbott20b}, which was obtained from the gravitational  wave signal by Bayesian method with the spectral EoS parametrizations~\cite{abbott18}.}The pressures from TM1 and NL3 sets around saturation density are larger than those from IUFSU and BigApple sets since their incompressibilities $K_0$ are about $270-280$ MeV. At the high-density region, the EoSs from NL3 and BigApple become stiffer since $g_3$ in these two sets are negative, which will produce smaller scalar fields and provide less attractive contributions to the EoSs. In DDRMF parameter sets, the EoSs have similar density-dependent behaviors. The DDV, DDVT, DDVTD, and TW99 sets generate the softer EoSs compared to other sets.

\begin{figure}[htb]
	\centering
	\includegraphics[scale=0.5]{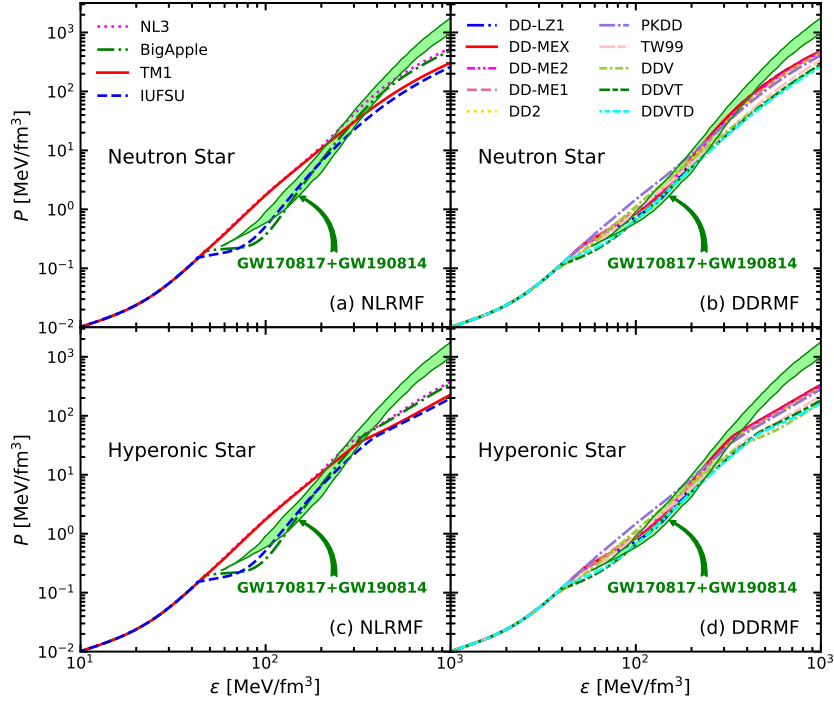}
	\small
	\caption{The pressure $P$ versus energy density $\varepsilon$ of neutron star matter and hyperonic star matter from NLRMF and DDRMF models. The joint constraints on EoS extracted from the GW170817 and GW190814 are shown as a shaded green band. Panels (a) and panel (b) for the neutrons star matter from the NLRMF and DDRMF models, respectively. Panels (c) and panel (d) for the hyperonic star matter from the NLRMF and DDRMF models, respectively. }\label{fig.1}
\end{figure}

After considering more conditions of beta equilibrium and charge neutrality about the hyperons, the EoSs of hyperonic star matter from the NLRMF  and  DDRMF models, where the coupling strengths between the mesons and hyperons are described as before, can be obtained in panel (c) and panel (d) of Fig.~\eqref{fig.1}. They almost become softer from $\varepsilon\sim 300$ MeV/fm$^{3}$ compared to the neutron star matter, due to the appearances of hyperons. In high-density region, all of them are below the joint constraints on the EoS from GW170817 and GW190814 events.

Similarly, in panel (a) and panel (b) of Fig.~\eqref{fig.2}, the pressures as functions of baryon density in neutron star matter from NLRMF and DDRMF models are given.  Furthermore, the speeds of sound in neutron star matter, $c_s$ with the unit of light speed are plotted in the insert. Since the NLRMF and DDRMF models are constructed based on the relativistic theory, the speeds of the sound from the RMF framework should be less than $1$ due to the causality. The NL3 set brings out the largest speed of the sound, while the $c^2_s/c^2$ from TM1 and IUFSU sets at high density approach $0.4$. It is noteworthy that $c^2_s/c^2$  is $1/3$ from the conformal limit of quark matter~\cite{malerran19,annala20}.
\begin{figure}[htb]
	\centering
	\includegraphics[scale=0.5]{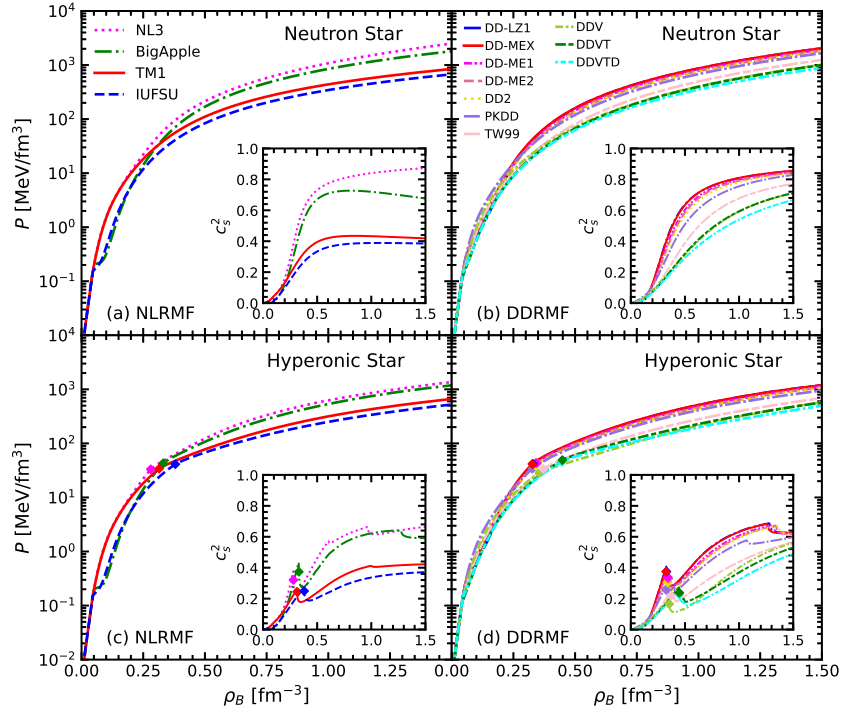}
	\caption{EoSs of neutron star and hyperonic star matter with different NLRMF and DDRMF models. The corresponding speeds of sound in units of the speed of light shown in the insert.  Panels (a) and panel (b) for the neutrons star matter from the NLRMF and DDRMF models, respectively. Panels (c) and panel (d) for the hyperonic star matter from the NLRMF and DDRMF models, respectively. }\label{fig.2}
\end{figure}

The pressures as functions of density in hyperonic star matter from NLRMF and DDRMF models are given in panel (c) and panel (d), respectively. The onset densities of the first hyperon are marked by the full discretized symbols, which are around $0.28-0.45$ fm$^{-3}$. The speed of sound of hyperonic star matter is not smooth as a function density anymore. The appearance of hyperon can sharply reduce the magnitude of the speed of sound, especially at first onset density. For the hard EoS, the $c^2_s$ becomes $0.6$ in hyperonic star matter from $0.8$ in neutron star matter at high-density region.

The onset densities of $\Lambda,~\Sigma$, and $\Xi$ hyperons in various NLRMF and DDRMF models are listed in Table.\eqref{table.5}. In general, the $\Lambda$ hyperon firstly arises around $2-3\rho_0$ due to its small mass and large attractive $\Lambda N$ potential. The most probable baryon of the second onset is the $\Sigma^-$ hyperon {for $\rho_B<4\rho_0$}, whose mass is very close to that of the $\Lambda$ hyperon. In a few parameter sets, such as NL3, PKDD, DDVT, and DDVTD, the second appearing hyperon is the $\Xi^-$ hyperon, which can bind with the nucleons to form the $\Xi^-$ hypernuclei. With the density increasing, the $\Xi^-$ hyperon usually emerges above $4\rho_0$ and $\Xi^0$ appears above $7\rho_0$. 
\begin{table}[htb]
	\small
	\centering
	\caption{Hyperon thresholds calculated with different RMF effective interactions for hyperonic matter. The unit of the density is fm$^{-3}$.}\label{table.5}
	\scalebox{0.80}{
		\begin{tabular}{lcccccccccccccccc}
			\hline\hline
			&$\rm 1^{st}$ ($\rho_{\rm th}$) ~~&$\rm 2^{nd}$ ($\rho_{\rm th}$) ~~&$\rm 3^{rd}$ ($\rho_{\rm th}$) ~~&$\rm 4^{th}$ ($\rho_{\rm th}$) \\
			\hline
			NL3          &$\Lambda$ (0.2804)  ~~&$\Xi^-$ (0.6078)  ~~&$\Xi^0$ (0.9723)  ~~&$\Sigma^-$(1.3545) \\
			BigApple     &$\Lambda$ (0.3310)  ~~&$\Sigma^-$ (0.4895)  ~~&$\Xi^-$ (0.6191)  ~~&$\Xi^0$ (1.2758) \\
			TM1          &$\Lambda$ (0.3146)  ~~&$\Sigma^-$ (0.9995)  ~~&$\Xi^-$ (1.0228)  ~~&     \\
			IUFSU        &$\Lambda$ (0.3800)  ~~&$\Sigma^-$ (0.5645)  ~~&                  ~~&     \\
			\hline\hline
			DD-LZ1        &$\Lambda$ (0.3294)  ~~&$\Sigma^-$ (0.4034)  ~~&$\Xi^-$ (0.6106)  ~~&$\Xi^0$ (1.2935)     \\
			DD-MEX        &$\Lambda$ (0.3264)  ~~&$\Sigma^-$ (0.3871)  ~~&$\Xi^-$ (0.5967)  ~~&$\Xi^0$ (1.2699)      \\
			DD-ME2        &$\Lambda$ (0.3402)  ~~&$\Sigma^-$ (0.4244)  ~~&$\Xi^-$ (0.4895)  ~~&$\Xi^0$ (1.3237)     \\
			DD-ME1        &$\Lambda$ (0.3466)  ~~&$\Sigma^-$ (0.4424)  ~~&$\Xi^-$ (0.4740)  ~~&$\Xi^0$ (1.3545)     \\
			DD2           &$\Lambda$ (0.3387)  ~~&$\Sigma^-$ (0.4147)  ~~&$\Xi^-$ (0.5699)  ~~&$\Xi^0$ (1.3733)     \\
			PKDD          &$\Lambda$ (0.3264)  ~~&$\Xi^-$ (0.4016)  ~~&$\Sigma^-$ (0.5126)  ~~&$\Xi^0$ (1.0759)      \\
			TW99          &$\Lambda$ (0.3696)  ~~&$\Sigma^-$ (0.4167)  ~~&$\Xi^-$ (0.7109)  ~~&$\Xi^0$ (1.7052)     \\
			DDV           &$\Lambda$ (0.3547)  ~~&$\Sigma^-$ (0.4850)  ~~&$\Xi^-$ (0.7723)  ~~&         \\
			DDVT          &$\Lambda$ (0.4465)  ~~&$\Xi^-$ (0.4941)  ~~&$\Sigma^-$ (0.6220)  ~~&        \\
			DDVTD         &$\Lambda$ (0.4465)  ~~&$\Xi^-$ (0.4963)  ~~&$\Sigma^-$ (0.6163)  ~~&          \\
			\hline\hline
	\end{tabular}}
\end{table}

\begin{figure*}[htb]
	\centering 
	\includegraphics[scale=0.38]{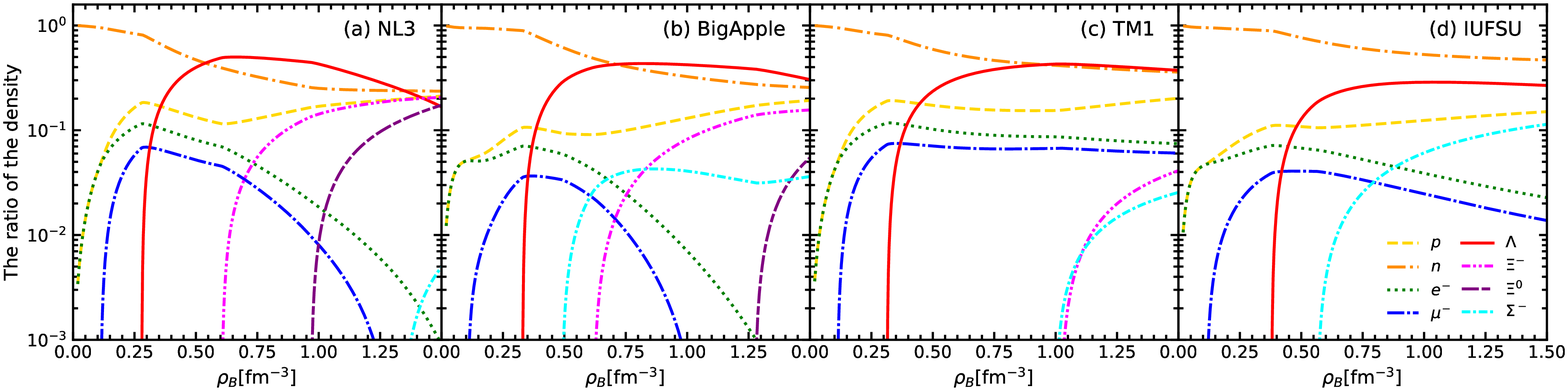}
	\small
	\caption{Particle fractions of baryons as a function of baryon number density with different NLRMF parameter sets. }\label{fig.3}
\end{figure*}
\begin{figure*}[htb]
	\centering
	\hspace*{-65pt}
	\includegraphics[scale=0.32]{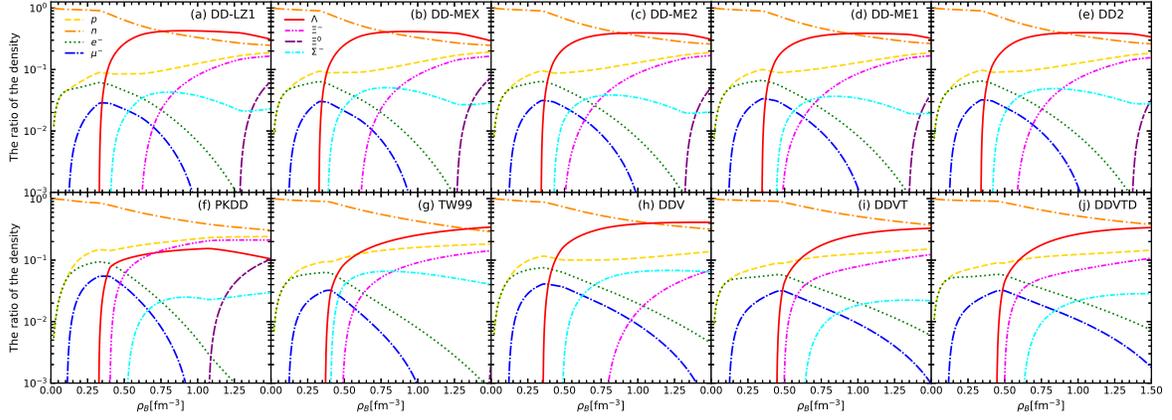}
	\small
	\caption{Particle fractions of baryons as a function of baryon number density with different DDRMF parameter sets. }\label{fig.4}
\end{figure*} 

The corresponding particle fractions of baryons as a function of baryon number density with different NLRMF and DDRMF parameter sets are shown in Fig.~\eqref{fig.3} and Fig.~\eqref{fig.4}, respectively. At a low-density region, the hyperonic matter is almost consisting of neutrons. The proton and electron fractions rapidly increase with density. When the chemical potential of the electron is larger than the free muon mass, the muon will arise. Above $2\rho_0$, the various hyperons appear in the hyperonic matter when they satisfy the chemical equilibrium conditions. At the high-density region, the fractions of various particles are strongly dependent on the $NN$ and $NY$ interactions. However, at all events, the fractions of $\Lambda$ hyperon will approach that of neutrons. In some cases, it can exceed those of neutrons.

After solving the TOV equation, the mass-radius relation of a static neutron star is obtained, where the EoSs of neutron star matter in the previous part are used. In Fig.~\eqref{fig.5}, the mass-radius ($M-R$) relations from NLRMF sets and DDRMF sets are shown in panel (a) and panel (b), respectively. The constraints from the observables of massive neutron stars, PSR J1614-2230 and PSR J034+0432 are also shown as the shaded bands. In 2019, the Neutron star Interior Composition Explorer (NICER) collaboration reported an accurate measurement of mass and radius of PSR J0030+0451, simultaneously. It may be a mass of $1.44_{-0.14}^{+0.15}M_\odot$ with a radius of $13.02_{-1.06}^{+1.24}$ km~\cite{miller19} and a mass of $1.34_{-0.16}^{+0.15}M_\odot$ with a radius of $12.71_{-1.19}^{+1.14}$ km~\cite{riley19} by two independent analysis groups. Recently, the radius of the pulsar PSR J0740+6620 with mass was reported by two independent groups based on NICER and X-ray Multi-Mirror (XMM-Newton) observations. The inferred radius of this massive NS is constrained to  $12.39_{-0.98}^{+1.30}$ km for the mass $2.072^{+0.0067}_{-0.066}M_{\odot}$ by Riley $et~ al.$~\cite{riley21}  and $13.7_{-1.5}^{+2.6}$ km for the mass $2.08\pm0.07M_\odot$ by Miller $et~al.$ at 68\% credible level~\cite{miller21}.  These constraints from NICER analyzed by Riley {\it et al.} are plotted in Fig.~\eqref{fig.5}. Meanwhile, the radius at $1.4M_{\odot}$ extracted from GW170817 is also shown~\cite{abbott18}.
\begin{figure}[htb]
	\centering
	\includegraphics[scale=0.5]{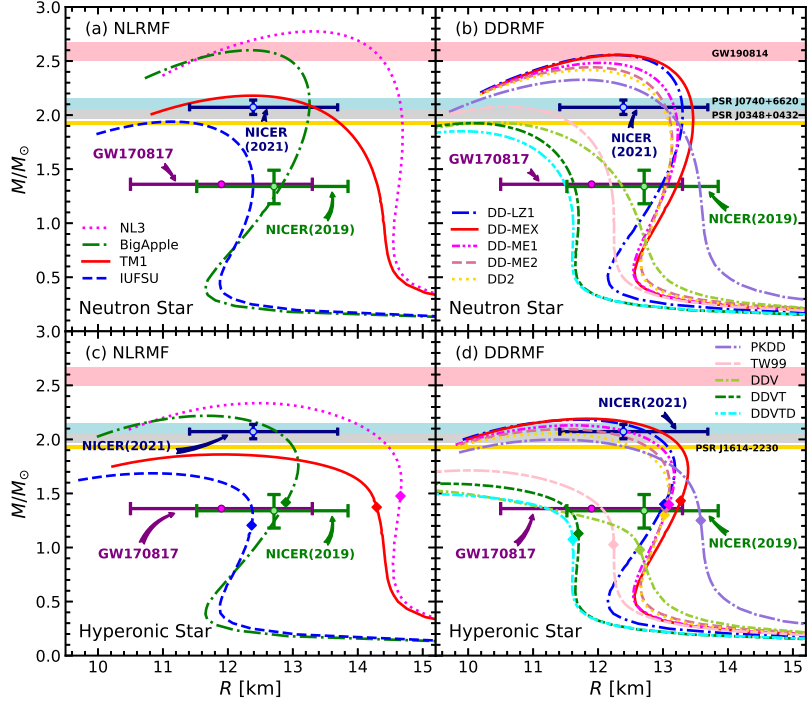}
	\caption{The neutron and hyperonic star masses as functions of  radius for NLRMF and DDRMF sets. Constraints from astronomical observables for the massive neutron star, NICER, and GW170817 are also shown. Panels (a) and panel (b) for the neutrons star matter from the NLRMF and DDRMF models, respectively. Panels (c) and panel (d) for the hyperonic star matter from the NLRMF and DDRMF models, respectively. }\label{fig.5}
\end{figure}

Among all parameter set used, NL3 predicts the heaviest neutron star mass ($2.77M_\odot$) due to its hard EoS, while the corresponding radius at $1.4M_\odot$ does not satisfy the measurements from GW170817 and NICER. {It can be found that the radii at $1.4M_\odot$ from NL3 and TM1 sets are much larger than those from other RMF sets. This is caused by their large slope of symmetry energy,  $L=110.6$ MeV from TM1 and $L=118.3$ MeV from NL3. In our previous works~\cite{ji19,hu20}, the extended TM1 and IUFSU parameter sets, which can generate different $L$ and keep the isoscalar properties of nuclear many-body systems, were applied to systematically study the symmetry energy effect on the neutron star. We found that there is a strong correlation between the $L$ and the radius of the neutron star at $1.4M_\odot$, while its influence on the maximum mass of the neutron star is very small. Furthermore, the tidal deformability, which is related to the radius of neutron also provides the constraints to $L$.}  From the present astrophysical observables, the slope of symmetry energy should be less than $80$ MeV. On the other hand, $L$ is also related to the neutron skin of the neutron-rich nuclei, such as $^{208}\rm{Pb}$. However, recent experimental data about the neutron skin of $^{208}\rm{Pb}$ from PREXII prefers the larger $L$~\cite{adhikari21,reed21,yue21}. This contradiction should be discussed in detail in the future. The softer EoSs from IUFSU, DDV, DDVT, and DDVTD cannot generate the $2M_\odot$ neutron stars and the radii at $1.4M_\odot$ from them are smaller compared to the other sets. BigApple, DD-LZ1, and DD-MEX sets can produce the neutron star heavier than $2.5M_\odot$, whose radii at $1.4M_\odot$ also accords with the constraints from gravitational wave and NICER. Therefore, we cannot exclude the possibility of the secondary in GW190814 as a neutron star~\cite{huang20}.  For the massive neutron stars above $2M_\odot$, from the stiffer EoSs, their central densities are less than $0.9$ fm$^{-3}$, while the softer EoSs may reach the central densities larger than $0.9$ fm$^{-3}$ for the lighter neutron stars.

The mass-radius ($M-R$) relations of hyperonic star from NLRMF and DDRMF parameter sets are shown in panel (c) and panel (d), respectively. The onset positions of the first hyperon in the relations are shown as the discretized symbols. After considering the strangeness degree of freedom, the maximum masses of the hyperonic star will reduce about $15\%\sim20\%$. In the NLRMF model, only the NL3 and BigApple parameter sets can support the existence of $2M_\odot$ compact star with hyperons, while in the DDRMF model, the DD-LZ1, DD-MEX, DD-ME2, DD-ME1, DD2, and PKDD sets generate the hyperonic star heavier than $2M_\odot$. Furthermore, the central densities of the hyperonic star become higher compared to the neutron star, all of which are above $5\rho_0$. The role of hyperons in the lower mass hyperonic star is strongly dependent on the threshold density of the first onset hyperon. The properties of a hyperonic star whose central density is below this threshold are identical to those of a neutron star. When the central density of the hyperonic star is larger than the threshold, the properties of hyperonic star will be influenced. For the softer EoSs, the lower mass neutron stars are more easily affected, because their central densities are much larger than those generated by the harder EoSs at the same neutron star mass. For example, the radii of the hyperonic stars at $1.4M_\odot$ from DDV, DDVT, and DDVTD decrease about $5\%$ compared to those of the neutron stars. 

In  Fig.~\eqref{fig.6}, the dimensionless tidal deformability, $\Lambda$, of neutron stars as a function of their masses from different NLRMF and DDRMF models are shown in panel (a) and panel (b), respectively. The tidal deformability represents the quadrupole deformation of a compact star in the external gravity field generated by its companion star, which is related to the mass, radius, and Love number of the star. From the gravitation wave of BNS merger in GW170817, it was extracted as $\Lambda_{1.4}=190^{+390}_{-120}$ at $1.4M_\odot$~\cite{abbott18}. It is found that the $\Lambda_{1.4}$ worked out by NLRMF  models are larger than the constraint of GW170817 since their radii are greater than $12$ km. For the massive neutron star, its tidal deformability is very small and close to $1$. The  $\Lambda_{1.4}$  from DDRMF models is separated into two types. The first type with the stiffer EoSs has the larger $\Lambda_{1.4}$ and heavier masses, whose $\Lambda_{1.4}$  are out the constraint of GW170817. The second one completely satisfies the constraints of GW170817 and has smaller radii from the softer EoSs. It was also shown the tidal deformability of the neutron star at $2.0 M_\odot$,  which is expected to be measured in the future gravitational wave events from the binary neutron-star merger. 

The dimensionless tidal deformabilities of hyperonic star are plotted in the panel (c) and panel (d) of Fig.~\eqref{fig.6}. For the harder EoSs, the hyperons only can influence the magnitudes of $\Lambda$ at the maximum star mass region, while they can reduce the dimensionless tidal deformability at $1.4M_\odot$ for the softer EoSs, such as TW99, DDV, DDVT, and DDVTD set, whose $\Lambda_{1.4}$ are closer to the constraints from GW170817.  Therefore, the compact stars in the GW170817 events may be the hyperonic stars.
\begin{figure}[htb]
	\centering
	\includegraphics[scale=0.5]{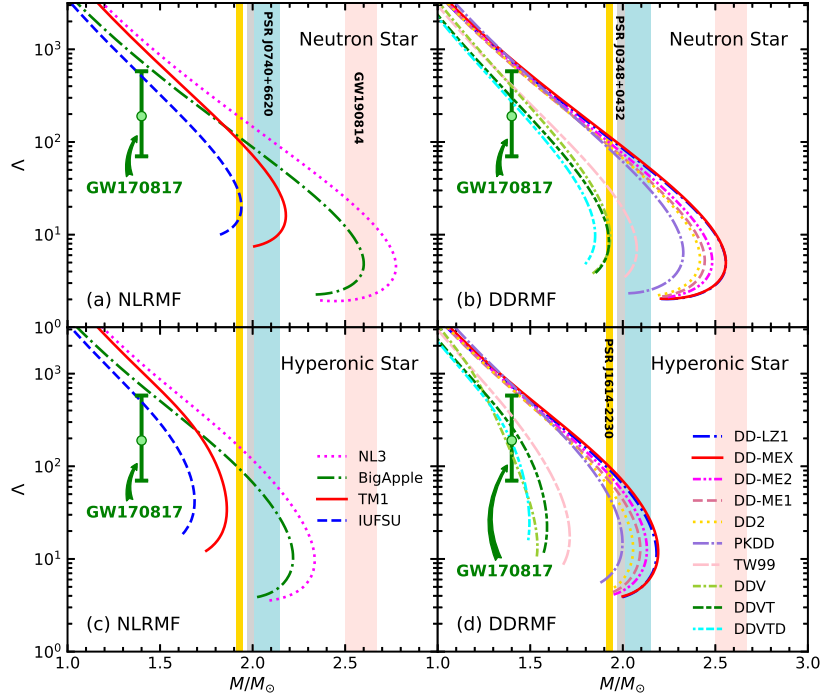}
	\caption{The dimensionless tidal deformability as a  function of star mass for NLRMF and DDRMF sets. The constraints from GW170817 event for tidal deformability is shown.  Panels (a) and panel (b) for the neutrons star matter from the NLRMF and DDRMF models, respectively. Panels (c) and panel (d) for the hyperonic star matter from the NLRMF and DDRMF models, respectively.}\label{fig.6}
\end{figure}

Finally, the properties of neutron star and hyperonic star, i.e., the maximum mass ($M_{\rm max}/M_{\odot}$), the corresponding radius ($R_{\rm max}$), the central density ($\rho_c$), the radius ($R_{1.4}$) and dimensionless tidal deformability ($\Lambda_{1.4}$) at $1.4M_\odot$ from present NLRMF and DDRMF models are listed in Table~\ref{table.6}. The maximum masses of neutron stars from these models are around $1.85\sim2.77M_\odot$, whose radii are from $9.93$ km to $13.32$ km. The central density of heavier neutron star is smaller. It is $0.66$ fm$^{-3}$ for NL3 set, while it becomes $1.28$ fm$^{-3}$ in DDVTD set. The radius at $1.4M_\odot$ from DDVTD set has the smallest value, $11.46$ km. Therefore, its dimensionless tidal deformability  at $1.4M_\odot$ is just $275$. 

Now, the maximum masses of hyperonic stars are $1.50\sim 2.34 M_\odot$, and the corresponding radii are in the range of $9.30\sim 12.51$ km, which are reduced compared to the cases without considering the strangeness degree of freedom. The central densities become larger compared to those of neutron stars. The smallest radius of the hyperonic star at $1.4M_\odot$ is $10.90$ km from DDV, whose dimensionless tidal deformability is just $136$. The threshold density of the first onset hyperon is also shown to demonstrate the influences of hyperon on the low mass hyperonic star. In general, the maximum mass of a hyperonic star can exceed $2M_\odot$ if the EoS is a hard type, whose maximum mass approaches $2.3M_\odot$ for a neutron star. Therefore, one solution to the ``hyperon puzzle" is to adopt the stiff EoS.  The softer EoS only can describe the hyperonic star whose mass is around $1.5M_\odot$.}

\begin{table*}[htb]
	\small
	\centering
	\caption{Neutron star and hyperonic star properties from various RMF models.}\label{table.6}
	\scalebox{0.65}{
		\begin{tabular}{l|cccccc|cccccc|cccc}
			\hline\hline
			&\multicolumn{6}{c|}{Neutron Star}& \multicolumn{5}{c}{Hyperonic Star}& \\
			\hline
			&$M_{\rm max}/M_{\odot}$ &$R_{\rm max}[\rm km]$ &$\rho_{c}[\rm fm^{-3}]$ &$R_{\rm 1.4}[\rm km]$ & $\rho_{1.4}[\rm fm^{-3}]$ &$\Lambda_{\rm 1.4}$ &$M_{\rm max}/M_{\odot}$ &$R_{\rm max}[\rm km]$ &$\rho_{c}[\rm fm^{-3}]$ &$R_{\rm 1.4}[\rm km]$ & $\rho_{1.4}[\rm fm^{-3}]$ &$\Lambda_{\rm 1.4}$  &$1^{\rm st}$ threshold [ fm$^{-3}$]\\
			\hline
			NL3          &2.7746  &13.3172  &0.6638  &14.6433  &0.2715  &1280  &2.3354  &12.5105  &0.8129   &14.6426 &0.2715  &1280  &0.2804   \\
			BigApple     &2.6005  &12.3611  &0.7540  &12.8745  &0.3295  &738   &2.2186  &11.6981  &0.8946  &12.8750  &0.3295  &738   &0.3310   \\
			TM1          &2.1797  &12.3769  &0.8510  &14.2775  &0.3200  &1050  &1.8608  &11.9255  &0.9736  &14.2775  &0.3218  &1050  &0.3146   \\
			IUFSU        &1.9394  &11.1682  &1.0170  &12.3865  &0.4331  &510   &1.6865  &10.8653  &1.1202  &12.3520  &0.4705  &498   &0.3800   \\
			\hline\hline                                                                 
			DD-LZ1       &2.5572  &12.2506  &0.7789  &13.0185  &0.3294  &729   &2.1824  &11.6999  &0.9113  &12.0185  &0.3294  &729   &0.3294  \\
			DD-MEX       &2.5568  &12.3347  &0.7706  &13.2510  &0.3228  &785   &2.1913  &11.8640  &0.8890  &13.2510  &0.3228  &785   &0.3264  \\
			DD-ME2       &2.4832  &12.0329  &0.8177  &13.0920  &0.3410  &716   &2.1303  &11.6399  &0.9296  &13.0920  &0.3410  &716   &0.3402  \\
			DD-ME1       &2.4429  &11.9085  &0.8358  &13.0580  &0.3512  &682   &2.0945  &11.5089  &0.9560  &13.0578  &0.3526  &681   &0.3466  \\
			DD2          &2.4171  &11.8520  &0.8481  &13.0638  &0.3528  &686   &2.0558  &11.3446  &0.9922  &13.0630  &0.3585  &685   &0.3387  \\
			PKDD         &2.3268  &11.7754  &0.8823  &13.5493  &0.3546  &758   &1.9983  &11.3789  &1.0188  &13.5400  &0.3642  &756   &0.3264  \\
			TW99         &2.0760  &10.6117  &1.0917  &12.1805  &0.4720  &409   &1.7135  &10.0044  &1.3466  &11.9880  &0.5710  &352   &0.3696  \\
			DDV          &1.9319  &10.3759  &1.1879  &12.3060  &0.5035  &395   &1.5387  &9.0109   &1.7317  &10.8990  &0.9538  &136   &0.3547  \\
			DDVT         &1.9253  &10.0846  &1.2245  &11.6058  &0.5458  &302   &1.5909  &9.6244   &1.4675  &11.4515  &0.6660  &266   &0.4465  \\
			DDVTD        &1.8507  &9.9294   &1.2789  &11.4615  &0.5790  &275   &1.4956  &9.3019   &1.6071  &10.9880  &0.8570  &182   &0.4465  \\
			\hline\hline
	\end{tabular}}
\end{table*}

\subsection{The correlations of hyperon coupling constants}\label{csc}
In recent work, Rong {\it et al.} ~\cite{rong21} found that the coupling ratio between the scalar meson and $\Lambda$ hyperon, $R_{\sigma \Lambda}$, has a strong correlation with that between the vector meson and $\Lambda$ hyperon, $R_{\omega \Lambda}$ in the available hypernuclei investigations by reproducing the single-$\Lambda$ binding energies from relevant experiments. This correlation is easily understood in the RMF framework since the single hyperon-nucleon potential given in Eq.~ \eqref{pot}, the single hyperon-nucleon potential is dependent on the scalar field and vector one. From the present observables of $\Lambda$ hypernuclei, we can extract that the $U^N_\Lambda \sim -30$ MeV at nuclear saturation density, $\rho_0$. On the other hand, the $\sigma_0$ and $\omega_0$ are solved in the symmetry nuclear matter, which are constants. Therefore, $R_{\sigma \Lambda}$ and $R_{\omega \Lambda}$ should satisfy the linear relation, when the $U^N_\Lambda$ is fixed in a RMF parameter set.

\begin{figure}[htb]
	\centering
	\includegraphics[scale=0.65]{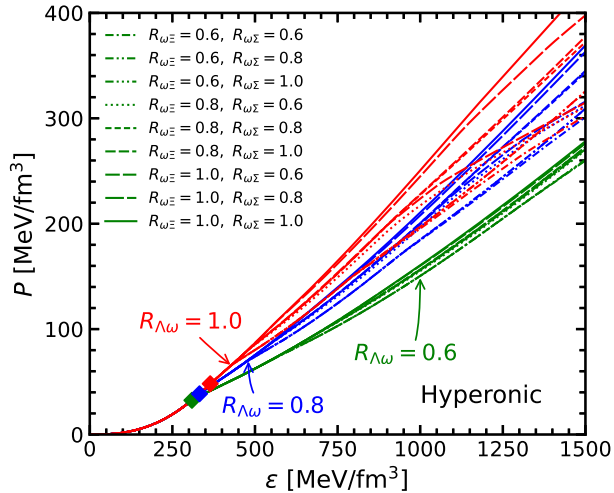}
	\small
	\caption{The EoSs of hyperonic star matter from TM1 models with different $R_{\omega\Lambda},~R_{\omega\Xi},~R_{\omega\Sigma}$. }\label{fig.11}
\end{figure}

In this subsection, the TM1 parameter set for $NN$ interaction will be adopted as an example to discuss the impact of the magnitudes of $R_{\sigma Y}$ and $R_{\omega Y}$ on the properties of hyperonic star under the constraints of $YN$ potential at nuclear saturation density, $U^N_Y(\rho_0)$.   From Eq.~\eqref{pot}, we can find a linear relation between the ratios $R_{\sigma Y}$ and $R_{\omega Y}$ ($Y=\Lambda,~\Sigma,~\Xi$) for different hyperons. In TM1 parameter set, the magnitudes of the scalar potential, $U_S=g_{\sigma N}\sigma^0$ and the vector potential, $U_V=g_{\omega N}\omega^0$ for nucleons are $342.521$ MeV and $274.085$ MeV at nuclear saturation density, respectively. With the empirical hyperon-nucleon potentials for $\Lambda,~\Sigma$ and $\Xi$ hyperons at nuclear saturation density,  $U_{\Lambda}^N=-30$ MeV, $U_{\Sigma}^N=+30$ MeV and $U_{\Xi}^N=-14$ MeV, the following relations {are} obtained,
\begin{align}
	R_{\omega\Lambda}=1.24969R_{\sigma\Lambda}-0.10946,\\
	R_{\omega\Sigma}=1.24969R_{\sigma\Sigma}+0.10946,\\
	R_{\omega\Xi}=1.24969R_{\sigma\Xi}-0.05108.
\end{align}
Here the strange mesons $\sigma^*$ and $\phi$ are not considered. Therefore, we can adjust the values of $R_{\omega Y}$ and generate the corresponding $R_{\sigma Y}$ simultaneously. To study the influences of $R_{\omega Y}$ on hyperonic star, $R_{\omega\Lambda},~R_{\omega\Xi},~R_{\omega\Sigma}=0.6,~0.8,~1.0$ are discussed, respectively. Therefore, there are $27$ combination cases.  The EoSs of hyperonic star matter from TM1 model with different $R_{\omega\Lambda},~R_{\omega\Xi},~R_{\omega\Sigma}$ are plotted in Fig.~\eqref{fig.11}. 
In general, the EoS becomes stiffer with the $R_{\omega\Lambda}$ increasing. For the $\Sigma$ and $\Xi$ coupling ratios, there are also similar behaviors. It demonstrates that the vector potential increases faster than the scalar one at the high-density region so that the EoS obtains more repulsive contributions from the vector meson.

\begin{figure}[htb]
	\centering
	\includegraphics[scale=0.65]{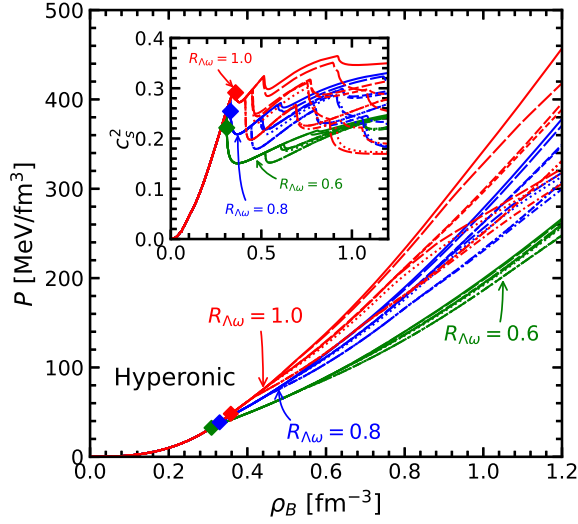}
	\small
	\caption{The pressure of hyperonic matter as a function of baryon density with different $R_{\omega\Lambda},~R_{\omega\Xi},~R_{\omega\Sigma}$. The corresponding speeds of sound in units of the speed of light shown in sub-figure. The threshold of the first hyperon is indicated by the filled diamonds. The meaning of the curves are same as those in Fig.~\eqref{fig.11}.}\label{fig.12}
\end{figure}

In Fig.~\eqref{fig.12}, the pressures and speeds of sound of hyperonic star matter as functions of baryon density with different $R_{\omega\Lambda},~R_{\omega\Xi},~R_{\omega\Sigma}$ are given. The thresholds of the first hyperon onset are symbolized by the filled diamonds. When the $R_{\omega\Lambda}$ is larger, the appearances of $\Lambda$ hyperons are later. Furthermore, the speeds of the sound of hyperonic matter are also strongly dependent on the magnitudes of $R_{\omega\Lambda},~R_{\omega\Xi},~R_{\omega\Sigma}$. The discontinuous places in these curves about the speeds of sound are generated by the onset of hyperons, which can reduce $c^2_s$. Therefore, the speed of sound of hyperonic star matter becomes very smaller, once the types of hyperons in hyperonic stars are raised more.	

\begin{figure}[htb]
	\centering
	\includegraphics[scale=0.5]{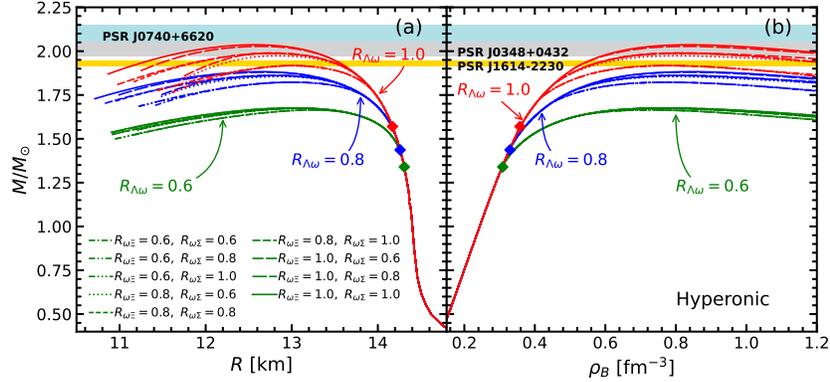}
	\small
	\caption{The hyperonic star masses as functions of  radius and the central baryon density for TM1 models with different $R_{\omega\Lambda},~R_{\omega\Xi},~R_{\omega\Sigma}$. The threshold of the first hyperon is indicated by the filled diamonds. }\label{fig.13}
\end{figure}

In Fig.~\eqref{fig.13}, the mass-radius ($M-R$) and mass-central density ($M-\rho_B$) relations from TM1 model with different $R_{\omega\Lambda},~R_{\omega\Xi},~R_{\omega\Sigma}$ are shown in panel (a) and (b), respectively.  The maximum mass of neutron star from the TM1 parameter set is $2.18 M_\odot$. When the hyperons are included, the maximum masses of the hyperonic star are reduced. They are just around  $1.67 M_\odot$ when $R_{\omega\Lambda}=0.6$. With the increment of  $R_{\omega\Lambda}$, the maximum mass of the hyperonic star becomes larger due to the stronger repulsive fields. It {is} around $2.0 M_\odot$ in the case of $R_{\omega\Lambda}=1.0$, which satisfies the constraints from the recent massive neutron star observables. The corresponding radii turn smaller and the central densities get larger.

Finally, the thresholds of hyperons and properties of hyperonic star with different $R_{\omega\Lambda},~R_{\omega\Xi},~R_{\omega\Sigma}$  are shown in Table.\eqref{table.9}.  In all of these 27 cases, the $\Lambda $ hyperon firstly appears in the hyperonic star.  Then, if $R_{\omega\Xi}>R_{\omega\Sigma}$, the $\Sigma^-$ hyperon will secondly emerge. Otherwise, the $\Xi^-$ hyperon appears. Furthermore, when $R_{\omega\Xi}<R_{\omega\Sigma}$, the $\Xi^0$ can more easily arises comparing to $\Sigma^0$ and $\Sigma^+$ hyperons. The maximum masses of hyperonic star and onset density of $\Lambda$ hyperon are strongly dependent on the $R_{\omega\Lambda}$ ratio. It can approach $2.04 M_\odot$ when $R_{\omega\Lambda}=1.0,~R_{\omega\Xi}=1.0,~R_{\omega\Sigma}=1.0$. The $\Lambda$ hyperon appears at $\rho_B=0.3089$ fm$^{-3}$ for $R_{\omega\Lambda}=0.6$, which is lower than the density of $1.4 M_\odot$ neutron star. Therefore, the radii and dimensionless tidal deformability at $1.4 M_\odot$, $R_{1.4}$ and $\Lambda_{1.4}$, are slightly changed as $12.2770$ km and $1055$ from the neutron star $12.2775$ km and $1050$. Therefore, the magnitudes of $R_{\omega\Lambda},~R_{\omega\Xi},~R_{\omega\Sigma}$ are significant to determine the maximum mass and corresponding radius of the hyperonic star, while it cannot by fixed very well through present hypernuclei experiments. 
\begin{table*}[htb]
	\footnotesize
	\centering
	\small
	\caption{Thresholds and hyperonic star properties from TM1 model with different $R_{\omega\Lambda},~R_{\omega\Xi},~R_{\omega\Sigma}$. The hyperons exist in the hyperonic star are given by bold.}\label{table.9}
	\scalebox{0.65}{
		\begin{tabular}{lcc|l|ccccccccccccc}
			\hline\hline 
			$R_{\omega\Lambda}$ &$R_{\omega\Xi}$  &$R_{\omega\Sigma}$  & \centering{Hyperon thresholds ($[\rm fm^{-3}]$) }
			&$M_{\rm max}$ ~&$R_{\rm max}[\rm km]$~&$\rho_{c}[\rm fm^{-3}]$~&$R_{1.4}[\rm km]$~&$\rho_{1.4}[\rm fm^{-3}]$~&$\Lambda_{1.4}$\\
			\hline
			0.6 &0.6 &0.6 &$\bm{\Lambda (0.3089),~\Xi^-(0.5245)},~\Sigma^- (0.9032),~\Xi^0(1.0710),~\Sigma^0(1.4447)$  &1.6645 &13.2465 &0.7251 &12.2770 &0.3255 &1055\\
			0.6 &0.6 &0.8 &$\bm{\Lambda(0.3089),~\Xi^-(0.5245)},~\Xi^0 (1.0759) $                                      &1.6645 &13.2465 &0.7251 &12.2770 &0.3255 &1055 \\
			0.6 &0.6 &1.0 &$\bm{\Lambda(0.3089),~\Xi^-(0.5245)},~\Xi^0(1.0759)$                                        &1.6645 &13.2465 &0.7251 &12.2770 &0.3255 &1055 \\
			0.6 &0.8 &0.6 &$\bm{\Lambda(0.3089),~\Sigma^-(0.6106),~\Xi^-(0.6394)},~\Xi^0 (1.3176),~\Sigma^0 (1.3237)$  &1.6733 &13.1347 &0.7456 &12.2770 &0.3255 &1055 \\
			0.6 &0.8 &0.8 &$\bm{\Lambda(0.3089),~\Xi^-(0.6306)},~\Sigma^-(1.2072),~\Xi^0(1.3924),~\Sigma^0 (1.8611)$   &1.6742 &13.1101 &0.7456 &12.2770 &0.3255 &1055 \\
			0.6 &0.8 &1.0 &$\bm{\Lambda(0.3089),~\Xi^-(0.6306)},~\Xi^0(1.3989)$                                        &1.6742 &13.1101 &0.7456 &12.2770 &0.3255 &1055 \\
			0.6 &1.0 &0.6 &$\bm{\Lambda(0.3089),~\Sigma^-(0.6106)},~\Sigma^0(1.2995),~\Sigma^+(1.4989) $               &1.6736 &13.1111 &0.7514 &12.2770 &0.3255 &1055 \\
			0.6 &1.0 &0.8 &$\bm{\Lambda(0.3089)},~\Sigma^-(0.7830),~\Xi^-(0.8546),~\Sigma^0(1.7530),~\Xi^0(1.8356)$    &1.6757 &13.0391 &0.7635 &12.2770 &0.3255 &1055 \\
			0.6 &1.0 &1.0 &$\bm{\Lambda(0.3089)},~\Xi^-(0.8237),~\Sigma^-(1.7052),~\Xi^0(1.8870)$                      &1.6757 &13.0391 &0.7635 &12.2770 &0.3255 &1055 \\
			\hline
			0.8 &0.6 &0.6 &$\bm{\Lambda(0.3294),~\Xi^-(0.4485),~\Sigma^-(0.6979)},~\Xi^0(0.8050),~\Sigma^0(1.0959)$    &1.8225 &13.0424 &0.7615 &12.2775 &0.3200 &1050 \\
			0.8 &0.6 &0.8 &$\bm{\Lambda(0.3294),~\Xi^- (0.4485)},~\Xi^0(0.8087)$                                       &1.8225 &13.0423 &0.7612 &12.2775 &0.3200 &1050 \\
			0.8 &0.6 &1.0 &$\bm{\Lambda(0.3294),~\Xi^-(0.4485)},~\Xi^0(0.8087)$                                        &1.8225 &13.0423 &0.7612 &12.2775 &0.3200 &1050 \\
			0.8 &0.8 &0.6 &$\bm{\Lambda(0.3294),~\Sigma^-(0.5009),~\Xi^-(0.5150)},~\Xi^0(0.9242),~\Sigma^0(0.9501)$    &1.8547 &12.8733 &0.7972 &12.2775 &0.3200 &1050 \\ 
			0.8 &0.8 &0.8 &$\bm{\Lambda(0.3294),~\Xi^-(0.5103)},~\Sigma^-(0.8429),~\Xi^0(0.9545),~\Sigma^0(1.3483)$    &1.8619 &12.7947 &0.8135 &12.2775 &0.3200 &1050 \\
			0.8 &0.8 &1.0 &$\bm{\Lambda(0.3294),~\Xi^-(0.5103)},~\Xi^0(0.9589)$                                        &1.8619 &12.7947 &0.8135 &12.2775 &0.3200 &1050 \\
			0.8 &1.0 &0.6 &$\bm{\Lambda(0.3294),~\Sigma^-(0.5009)},~\Sigma^0(0.9200),~\Sigma^+(1.0661),~\Xi^-(1.1906)$ &1.8603 &12.8329 &0.8026 &12.2775 &0.3200 &1050 \\
			0.8 &1.0 &0.8 &$\bm{\Lambda(0.3294),~\Sigma^-(0.5967),~\Xi^-(0.6220)},~\Sigma^0(1.1906),~\Xi^0(1.2128)$    &1.8799 &12.6868 &0.8316 &12.2775 &0.3200 &1050 \\
			0.8 &1.0 &1.0 &$\bm{\Lambda(0.3294),~\Xi^-(0.6135)},~\Sigma^-(1.1163),~\Xi^0(1.2699),~\Sigma^0(1.8870)$    &1.8828 &12.6492 &0.8371 &12.2775 &0.3200 &1050 \\
			\hline
			1.0 &0.6 &0.6 &$\bm{\Lambda(0.3579),~\Xi^-(0.4186),~\Sigma^-(0.6050),~\Xi^0(0.6947)},~\Sigma^0(0.9589)$    &1.9170 &13.0352 &0.7661 &12.2775 &0.3200 &1050 \\
			1.0 &0.6 &0.8 &$\bm{\Lambda(0.3579),~\Xi^-(0.4128),~\Xi^0(0.6979)},~\Sigma^-(1.4852),~\Sigma^0(1.5409)$    &1.9174 &12.9932 &0.7795 &12.2775 &0.3200 &1050 \\
			1.0 &0.6 &1.0 &$\bm{\Lambda(0.3579),~\Xi^-(0.4128),~\Xi^0 (0.6979)}$                                       &1.9174 &12.9932 &0.7795 &12.2775 &0.3200 &1050 \\
			1.0 &0.8 &0.6 &$\bm{\Lambda(0.3579),~\Sigma^-(0.4506),~\Xi^-(0.4590),~\Xi^0(0.7617),~\Sigma^0(0.8013)}$    &1.9736 &12.8272 &0.8047 &12.2775 &0.3200 &1050 \\
			1.0 &0.8 &0.8 &$\bm{\Lambda(0.3579),~\Xi^-(0.4548),~\Sigma^-(0.6947),~\Xi^0(0.7759)},~\Sigma^0(1.1061)$    &1.9878 &12.7682 &0.8093 &12.2775 &0.3200 &1050 \\
			1.0 &0.8 &1.0 &$\bm{\Lambda(0.3579),~\Xi^-(0.4548),~\Xi^0(0.7759)}$                                        &1.9879 &12.7682 &0.8093 &12.2775 &0.3200 &1050 \\
			1.0 &1.0 &0.6 &$\bm{\Lambda(0.3579),~\Sigma^-(0.4506),~\Xi^-(0.6726),~\Sigma^0(0.7617)},~\Sigma^+(0.8826)$ &1.9898 &12.7706 &0.8146 &12.2775 &0.3200 &1050 \\
			1.0 &1.0 &0.8 &$\bm{\Lambda(0.3579),~\Sigma^-(0.5126),~\Xi^-(0.5221)},~\Xi^0(0.9074),~\Sigma^0(0.9328)$    &2.0275 &12.6470 &0.8254 &12.2775 &0.3200 &1050 \\
			1.0 &1.0 &1.0 &$\bm{\Lambda(0.3579),~\Xi^-(0.5197)},~\Sigma^-(0.8546),~\Xi^0(0.9328),~\Sigma^0(1.4447)$    &2.0363 &12.5920 &0.8327 &12.2775 &0.3200 &1050 \\
			\hline\hline
	\end{tabular}}
\end{table*}

\section{Summary}
The neutron star consisting of nucleons and leptons, and the hyperonic star considering additional contributions from hyperons were reviewed in the relativistic mean-field (RMF) model due to recent rapid achievements about astronomical observations on the compact star. The theoretical frameworks of the two types of RMF models, i.e. non-linear RMF (NLRMF) and density-dependent RMF (DDRMF) were shown in detail to describe the infinite nuclear matter system. Several conventional NLRMF parameter sets, (NL3, BigApple, TM1, IUFSU), and the DDRMF parameter sets (DD-LZ1, DD-MEX, DD-ME2, DD-ME1, DD2, PKDD, TW99, DDV, DDVT, DDVTD) were adopted to calculate the properties of the neutron star and hyperonic star, which were created by reproducing the ground-state properties of several finite nuclei with different considerations.

The equations of state (EoSs) of neutron star matter from these parameterizations at high-density region are separated into the softer type (IUFSU, DDV, DDVT, DDVTD) and stiffer one (NL3, BigApple, TM1, DD-MEX, DD-ME2, DD-ME1, DD2, PKDD, TW99). The maximum masses of neutron stars generated by the softer EoSs cannot approach $2.0 M_\odot$, which did not satisfy the constraints from the massive neutron star observables. However, the radii of the corresponding neutron star are relatively small so that their dimensionless tidal deformability at $1.4 M_\odot$ are around $275\sim510$ which are in accord with the value extracted from the GW170817 event. Meanwhile, the harder EoS can lead to a very massive neutron star. The maximum masses are $2.7746 M_\odot$ and $2.5572 M_\odot$ from NL3 and DD-LZ1 sets, respectively, which implies that the secondary object in GW190814 may be a neutron star. In addition, the radius of the neutron star at $1.4 M_\odot$ has a strong correlation with the slope of symmetry energy, $L$.

The baryon-baryon interaction plays an essential role in the hyperonic star matter, which is extracted from the experimental data of the hypernuclei. The meson-hyperon coupling strengths in RMF parameter sets were generated by the empirical hyperon-nucleon potential in symmetric nuclear matter at nuclear saturation density. The strangeness scalar and vector mesons were introduced to consider the $\Lambda-\Lambda$ potential in hyperonic star with the bond energies of double $\Lambda$ hypernuclei. The hyperonic star matter becomes softer compared to the neutron star matter. The onset densities of the first hyperon were around $2\rho_0\sim3\rho_0$ in present RMF models. The hyperon was raised earlier in the stiffer EoS.  The appearance of hyperon can reduce the speed of the sound of the hyperonic star matter. The maximum mass of the hyperonic star is larger than $2M_\odot$ for the stiffer RMF parameter sets.  In addition, the hyperons influence the properties of the hyperonic star in the low-mass region from softer EoS since its central density is very higher. Therefore, dimensionless tidal deformability at $1.4 M_\odot$ will get smaller and be closer to the constraints from GW170817.

Finally, the magnitudes of the coupling strengths between scalar and vector mesons, and hyperons were discussed in the TM1 parameter set. When the single hyperon-nucleon potential in symmetric nuclear matter at nuclear saturation density was fixed at a  RMF parameter set, the coupling constant between scalar meson and hyperon $R_{\sigma Y}$ has the obvious linear correlations with that between vector meson and hyperon $R_{\omega Y}$. With present hypernuclei experimental data, it is difficult to completely determine the magnitudes of  $R_{\sigma Y}$ and  $R_{\omega Y}$. The linear correlations were applied to investigate the effect of $R_{\omega Y}$ strength on the hyperonic star. It was found that the maximum mass of a hyperonic star can arrive at $2M_\odot$ when  $R_{\omega Y}=1.0$, while from the conventional quark counting rules, it is just around $1.7 M_\odot$. Furthermore, the onset densities of various hyperons in hyperonic stars were also strongly dependent on the magnitudes of $R_{\omega Y}$. The speeds of sound of hyperonic star matter will largely reduce when the types of appearance hyperons increase.

The strangeness degree of freedom can largely affect the properties of hyperonic stars in the RMF framework and reduce the maximum mass of the hyperonic star. However, the ``hyperon puzzle" was completely solved with the stiffer EoS generated by many available RMF parameter sets. The DDRMF models more easily generate the massive hyperonic star due to their density dependence of coupling constants. On the other hand, it is also obtained by increasing the coupling constants between the vector meson and hyperons. The gravitational wave provides a good manner to study the structure of the compact star. It is hoped that there will be special signal from gravitational waves, which can distinguish the existence of hyperons in the hyperonic star. More experiments about the hypernuclei are expected to obtain more information about the hyperon-nucleon potential, which will greatly promote the investigations about the hyperonic star. 

\section*{Acknowledgments} 
This work was supported in part by the National Natural Science Foundation of China (Grants  Nos. 11775119 and 12175109) and  the Natural Science Foundation of Tianjin.

\bibliographystyle{apsrev}
\bibliography{Reference}

\end{document}